\documentclass[aip,cha,reprint,citeautoscript]{revtex4-1}
\pdfoutput=1

\usepackage{geometry}
\usepackage[utf8]{inputenc}

\usepackage[tight]{subfigure}
\usepackage{graphicx}
\graphicspath{{img/}}
\DeclareGraphicsExtensions{.pdf,.png}
\setkeys{Gin}{height=30mm,keepaspectratio} % default height

\usepackage[textsize=scriptsize,colorinlistoftodos]{todonotes}

\usepackage{paralist}

\usepackage{amsmath,amssymb,amsthm}
\newcommand{\abs}[1]{ \ensuremath{\left\lvert{#1}\right\rvert}}

\theoremstyle{definition}
\newtheorem*{definition*}{Definition}%[section]

\DeclareMathOperator{\erfc}{erfc}
\DeclareMathOperator{\ftbe}{FTBE}

\newcommand{\arefmodel}{Aref Blinking Vortex flow}
\newcommand{\subbraid}{subbraid}  % prefer subbraid to sub-braid

\usepackage{hyperref}
\hypersetup{
  bookmarksopen=True,
  bookmarksnumbered=True
}

\begin{document}
  \affiliation{Department of Mathematics,
    University of Wisconsin, Madison, WI, USA}
  \author{Marko Budi\v{s}i\'{c}}
  \email{marko@math.wisc.edu}
  \author{Jean-Luc Thiffeault}
  \email{jeanluc@math.wisc.edu}
  %\date{\texttt{\today~DRAFT}}
  \title{Finite-Time Braiding Exponents}
  \begin{abstract}
    Topological entropy is a common measure of the rate of mixing in a flow.
    It can be computed by partition methods, or by estimating the growth rate
    of material lines or other material elements.  This requires detailed
    knowledge of the velocity field, which is not always available, such as
    when we only know a few particle trajectories (ocean float data, for
    example).  We propose an alternative approximation to topological entropy,
    applicable to two-dimensional flows, which uses only a finite number of
    trajectories as input data. To represent these sparse data sets, we use
    \emph{braids}, algebraic objects that record how strands, i.e.,
    trajectories, exchange positions with respect to a projection
    axis. Material curves advected by the flow are represented as simplified
    \emph{loop coordinates}. The exponential rate at which a braid deforms
    loops over a finite time interval as the strands exchange places is the
    Finite-Time Braiding Exponent (FTBE) and serves as a proxy for topological
    entropy of the two-dimensional flow. We demonstrate that FTBEs are robust
    with respect to the value of numerical time step, details of braid
    representation, and choice of initial conditions inside the mixing
    region. We also explore how closely the FTBEs approximate topological
    entropy depending on the number and length of trajectories used.
  \end{abstract}

\maketitle{}

\begin{quotation}
  In geophysical flows and many other applications it is important to know
  where things can go, and where they come from.  This is the study of
  transport and its cousin, mixing.  Modern methods are most powerful when we
  know the flow perfectly through its velocity field.  But when the data comes
  from, say, ocean floats, it is very sparse and cannot completely
  characterize transport.  We use tools from the mathematical branch of
  topological dynamics to tease out as much information as possible about
  mixing in the flow, in particular the degree of ``entanglement'' that a set of
  trajectories achieves.  We call this measure the Finite-Time Braiding
  Exponent.
\end{quotation}

\section{Introduction}
\label{sec:introduction}

A common approach to studying kinematics of physical fluid flows is through
their representation as dynamical systems, either as differential equations
that govern paths of advected particles, or as maps that connect initial and
terminal locations of the particles. When the dynamics are
deterministic, the complexity of the flow is connected to the chaotic nature of
particle trajectories, a phenomenon called
\emph{chaotic advection}~\cite{Aref1984,Ottino1989}.

In dynamical systems, and especially ergodic theory, the amount of complexity
present in the system is quantified by various notions of
entropy.~\cite{Young2003a} Informally, positive entropy in the system is the
indicator of sustained complex behavior of trajectories. Formally, there are
several non-equivalent notions of entropy that can be used to characterize a
dynamical system. The two most common entropies in deterministic dynamical
systems are metric (Kolmogorov--Sinai) entropy and topological entropy.

Both metric and topological entropies can be described by considering a
setting in which two trajectories of length \(T\) can be distinguished only if
they are further than some resolution \(\varepsilon\) apart. If we count the
number of \(\varepsilon\)-distinct trajectories of length \(T\), we expect
their number to grow as \(T \to \infty\). If the growth is asymptotically
exponential \(\sim e^{h T}\), then the rate \(h\) is the entropy of the
flow. The crucial difference between topological and metric entropy is in the
way that \(\varepsilon\)-distinct orbits are counted: topological entropy
counts \emph{all} \(\varepsilon\)-distinct orbits, while metric entropy counts
only \emph{typical} \(\varepsilon\)-distinct orbits, where ``typical'' allows
for ignoring a set of orbits that is of zero measure with respect to an
invariant measure\cite{Young2003a}. This means that each invariant measure has an associated
metric entropy, and topological entropy is their supremum.  An
immediate practical consequence is that topological entropy provides an upper
bound for metric entropy. For a further theoretical overview, see
ref.~\onlinecite{Young2003a}; for applications, see
refs.~\onlinecite{Bollt2013,Haszpra2013a}.

Counting distinct orbits to estimate either of the entropies is feasible only
for certain classes of dynamical systems.\cite{Bowen1978} However, topological
entropy can be estimated by discretizing the advection operator. This approach
has been used both for analysis\cite{Froyland2001,Froyland2012a} and
synthesis\cite{DAlessandro1999a} of dynamics.

Alternatively, metric entropies can be estimated using Lyapunov exponents,
which measure the rate of separation of particles as they are advected by the
flow.  Lyapunov exponents are the mean of local
deformation rates along trajectories of the flow. Local measurements of
deformation --- Lyapunov exponents --- and global measurements of complexity
--- entropies --- are connected by the Margulis--Ruelle inequality which states
that the spatial mean of positive Lyapunov exponents with respect to the
invariant measure is an upper bound for the metric entropy with respect to that
measure.  The Pesin formula strengthens the inequality to an equality for
those invariant measures that are non-singular with respect to the standard
volume measure.~\cite{Katok1980,Young2003a}

In applied problems, trajectories are almost always observed over finite
times, and thus definitions of entropies do not strictly apply. Nevertheless,
rates of deformation are used as indicators of complexity in geophysical
dynamics, in addition to statistical quantities such as relative dispersion
rates~\cite{Waugh2012} and effective diffusivity.~\cite{Marshall2006} In
particular, the amount of mixing in a region is commonly measured by
exponential rates of separation of fluid trajectories, either over finite time
horizons (Finite-Time Lyapunov Exponents,~\cite{Pierrehumbert1991} or FTLEs) or
up to finite scales (Finite-Size Lyapunov Exponents\cite{Aurell1997,
  Artale1997}, or FSLEs). Computation of FTLE and FSLE fields gained
prominence in studying barriers to material transport in fluid
flows~\cite{Joseph2002, dOvidio2004, Farnetani2003}, as ridges and troughs of
FTLE/FSLE fields were used as proxies\cite{Haller2002, Haller2001a,
  Haller2000, Shadden2005, Haller2011} to Lagrangian Coherent Structures (for
an introduction to the topic see ref.~\onlinecite{Haller2015}). Despite some
issues in using FTLE and FSLE for detection of Lagrangian Coherent
Structures,~\cite{BozorgMagham2013, Bollt2013, Karrasch2013} these techniques
remain popular in analysis of geophysical fluid flows~\cite{Samelson2013}.

In experimental and numerical fluid flows, Lyapunov exponents might be
difficult to compute efficiently and reliably as they require knowledge of the
gradient tensor of the velocity field, and precise temporal resolution of
data\cite{LaCasce2008}. As an alternative, topological entropy of
two-dimensional flows governed by ODEs can be estimated by measuring growth of
material curves, as shown by Newhouse and Pignataro~\cite{Newhouse1993}. A
material curve is a curve of initial conditions advected as a set by the
dynamical system. Material curves are expected to grow exponentially fast in
chaotic flows; the largest exponential rate, over all material curves, is
equal to the topological entropy in smooth flows, or else it provides
a lower bound.  In practice the advection of
material curves requires tracking an exponentially-growing number of
trajectories, which can exceed a computer's memory.

We work at the other extreme of available data, assuming that a sparse set of
trajectories is the only information known about the flow. While this regime
seems restrictive, such data sets are common in physical oceanography,
comprising trajectories of floats and drifters passively advected by
currents\cite{Mariano2002,LaCasce2008}. Other examples include granular
media~\cite{Puckett2012}, crowds of people~\cite{Ali2013}, and animal
flocks~\cite{Caussin2015, Topaz2014}, all of which can behave similarly to
particles advected by flows, without being driven by physically-observable
vector fields. Even when detailed models of flow velocities are available, it
might be preferable to compute coarse estimates based on several trajectories
when speed of analysis is important. In all these examples, topological
approaches have proved useful to measure complexity of two-dimensional
dynamics.

The topological theory for measuring rates of deformation is intuitively
similar to direct material line advection. However, instead of a detailed
bookkeeping of material curves and particle trajectories, the data is
simplified drastically using two symbolic representations: braids and
loops. The trajectories are represented as braids: time-ordered sequences of
symbols that encode the manner in which trajectories exchange their order
along a chosen axis.~\cite{Boyland2000, Thiffeault2005} Material lines are
represented by topological loops: closed ``rubber bands'' that wrap around
trajectories.~\cite{Thiffeault2010} Both braids and loops can be economically
represented using symbols: generators for braids, and loop coordinates for
loops.  Acting on loops with braids amounts to studying tight material curves
that enclose a set of trajectories.  These curves grow as the trajectories
evolve in time.  Because all this is done using symbolic representations,
estimates of loop growth can be computed very efficiently compared to the full
model.

In applied dynamical systems, braids were first used to represent dynamical
evolution of points that lie on periodic orbits.~\cite{Boyland1994,
  Boyland2000, Finn2006, Gouillart2006, Thiffeault2006} When trajectories are
periodic, recording them over one common interval and representing them using
braid generators results in an element of the braid group. The entropy of
braids can be computed very precisely, sometimes even analytically (or at
least in terms of the largest root of a known polynomial).  Braids
conveniently label classes of continuous mappings of two-dimensional
domains. As topological entropy of the flow is associated with its flow map,
it can be shown that entropy of a braid of a periodic orbit in such a flow is
a lower bound for topological entropy of the flow.~\cite{Handel1985,
  Boyland2000, Gouillart2006, Thiffeault2006, %
  Tumasz2012, Tumasz2012a, Tumasz2013}

When measured trajectories are not periodic,
a true topological representation would involve an infinite braid.
Nevertheless, we can still estimate exponential rates of stretching of loops
using braids of finite trajectories, similar to calculations given in
ref.~\onlinecite{Thiffeault2010}. We use the name \emph{Finite-Time Braiding
  Exponent} for the proposed complexity measure, as it intuitively corresponds
to Finite-Time Lyapunov Exponent calculation. However, FTBEs differ from FTLEs
in that the latter is only a local measure of deformation around a single
trajectory.

Computing FTBEs amounts to finding the maximal rate of deformation that a loop
can experience under the action of the braid; however, not all loops deform at
the same rate. Allshouse and Thiffeault~\cite{Allshouse2012} showed that
slow-growing loops often encircle coherent structure boundaries. In flows with
sparse trajectory data, a more detailed understanding of loop deformation by
braids of trajectories is thus relevant for both identification of coherent
structures and measurement of mixing.

Other than several graphs in refs.~\onlinecite{Thiffeault2005,Thiffeault2010}
there are no results about how FTBEs depend on the way in which data is
collected, or how tightly they might approximate the topological entropy of
the flow.  A theorem in this direction would be invaluable, but in its absence
we explore how FTBEs depend on \begin{inparaenum}[1.]
\item the number of trajectories in the data set
  (Section~\ref{sec:dependence-on-nstrands});
\item the location of initial conditions for trajectories; and
\item the length of trajectories (both in Section~\ref{sec:init-cond-length}) .
\end{inparaenum}
Numerical computations with the \arefmodel{} indicate that the spatial
variance of FTBEs decays with both increased number and length of
trajectories.  The mean FTBE increases with the number of trajectories until a
saturation is achieved.  We propose models for this growth in
Section~\ref{sec:extrapolation}, as well as compare the saturation bound with
other estimates of topological entropy.

\section{Mathematical background}
\label{sec:definition}

As mentioned in the introduction, we assume that the only data known about a
dynamical system on a two-dimensional domain is a finite number of
trajectories of equal, finite lengths. In this section, we describe how to
represent the trajectories as braids and how to calculate Finite-Time Braiding
Exponents from the braid. Finally, we briefly describe \texttt{braidlab}\cite{Thiffeault2014v3}, a
publicly-available MATLAB toolbox implementing the calculation of braids and
FTBEs.

\subsection{Representation of braids and loops}
\label{sec:braids-and-loops}

Our dynamics will be assumed to be defined on a closed unit disk \(\mathbb D\)
in a plane. Since the braid theory approach uses only topological arguments,
any other domain homeomorphic to a disk, i.e., that can be continuously
deformed to a disk, will require no special treatment.

A \emph{physical braid} (or geometric braid) is a collection of finitely-many
planar continuous trajectories on a (bounded) time interval \(I = [t_{0},
t_{0}+T]\). When they are thought of as curves in the extended
three-dimensional space \(\mathbb{D} \times I\), trajectories are also called
\emph{strands} (or strings). At any time \(t \in I\), strands evaluate to a
set of \(n\) distinct points, often called \emph{punctures}, which move around
in time.

Constructing a braid from a physical braid is a method of discarding
information about geometry (e.g., distances between strands), while retaining
information about topology (e.g., relative locations of the strands).  We keep
only those time instances, called \emph{crossings}, in which there is a
substantial change in the positions of strands. Crossings are defined with
respect to a projection axis --- a line passing through the origin of the
domain at an angle \(\alpha\).~\cite{Thiffeault2005, Thiffeault2010} Strands
are ordered and indexed \(1,2,\dots,n\) according to their projection onto the
axis.  As the punctures move in time, they will exchange their indices when
they cross.

A \emph{braid} \(\mathbf{b}\) is a sequence of symbols \(\mathbf{b} =
b_{1}b_{2}\cdots\) where each symbol corresponds to a crossing in a physical
braid. The symbols are taken from the set of \(2(n-1)\) braid generators
\(\{\sigma_{1}^{\pm1}, \sigma_{2}^{\pm1}, \dots, \sigma_{(n-1)}^{\pm1}\}\).
Each symbol \(\sigma_{i}^{\pm 1}\) corresponds to a crossing of strands \(i\)
and \(i+1\), either clockwise \((+1)\) or anticlockwise
\((-1)\).~\cite{Thiffeault2005}

\emph{Loops} are counterparts of advected material lines, in the same way as
braids are counterparts of trajectories. To specify a loop in the disk, draw
one or more closed curves that do not intersect themselves, each other, the
boundary, nor pass through one of the punctures. In the topological picture,
curves can be continuously deformed into each other without passing through
one of the punctures. This allows us to ``tighten'' each loop around the
punctures it encloses. Loops that can be tightened to a single puncture or
expanded to the disk boundary are never affected by the motion of punctures;
we do not consider those loops.  The remaining (\emph{essential}) loops are
deformed by the motion as if they were rubber bands caught on the punctures.

We encode loops with a particular integer coordinate system, introduced by
Dynnikov~\cite{Dynnikov2002} and explained in detail in
refs.~\onlinecite{Hall2009,Thiffeault2010}. Dynnikov coordinates provide a
one-to-one and onto correspondence between essential loops and vectors in
\(\mathbb{Z}^{2n-4}\). Each braid generator \(\sigma_{i}^{\pm1}\) acts on
loops via a piecewise-linear function \(\sigma_{i}^{\pm1}:\mathbb{Z}^{2n-4}
\to \mathbb{Z}^{2n-4}\), while the action of full braid is formed by
composition of such functions.  The functions are given in
ref.~\onlinecite{Hall2009,Thiffeault2010}.

Using braids and Dynnikov coordinates we represent the dynamics of material
lines by integer-valued functions. Given a braid \(\mathbf{b}\) and a vector
\(v \in \mathbb{Z}^{2n-4}\) encoding a loop, we will use multiplication
\(\mathbf{b}\cdot v\) (or \(\mathbf{b}v\)) to denote the nonlinear action of
\(\mathbf{b}\) on \(v\).

\subsection{Finite-Time Braiding Exponents}
\label{sec:ftbe}

The \emph{word length} \(L(\mathbf{b})\) of a braid \(\mathbf{b} =
b_{1}b_{2}\cdots b_{k}\) is~\(k\). While trivial to compute,\footnote{The
  \emph{minimum word length} is much more difficult to compute (see
  ref.~\protect\onlinecite{Paterson1991, Bangert2002}).  Here we simply count
  the number of crossings and do not attempt to simplify the braid.} the word
length does not capture topological complexity very well. For example, imagine
a set of strands that are sparsely placed on the domain, and dynamics that
only ``wiggles'' them slightly. With certain choices of projection angle
\(\alpha\), wiggles will produce plenty of crossings among the strands, with
no topological complexity whatsoever.~\cite{Thiffeault2005}

Braid entropy is a measure of braid complexity derived from topological
entropy of continuous maps, specifically pseudo-Anosov
maps~\cite{Fathi1979}. In topological surface dynamics, braids are used as
labels for classes of continuous maps that can be continuously deformed into
each other (homotopy-equivalent homeomorphisms, see
refs.~\onlinecite{Birman1975,Farb2012,Boyland1994}). The topological
entropy of a homeomorphism is directly connected to the maximal
\emph{asymptotic} exponential rate of stretching that a curve on a surface
will experience under repeated action of the homeomorphism.~\cite{Bowen1978,
  Fathi1979, Franks1988, Newhouse1988, Newhouse1993, Moussafir2006} The
topological entropy of a braid \(h(\mathbf{b})\) is then the \emph{minimal}
entropy of a homeomorphism out of the entire class of homeomorphisms labeled
by the braid \(\mathbf{b}\).

\begin{figure}%[h!]
  \centering
  \subfigure[{} \(\ell_{E}\) \label{fig:loop-e} ]{\includegraphics[width=.3\textwidth]{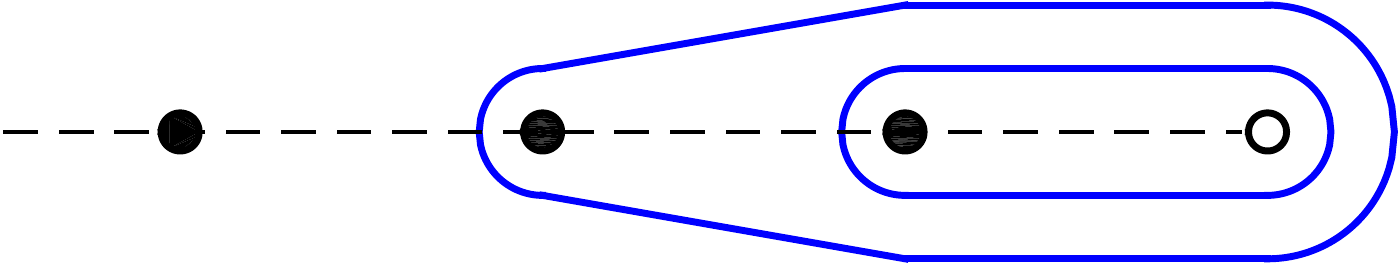}}\\
  \subfigure[{} \(\sigma_{1}\sigma_{1} \cdot \ell_{E}\) \label{fig:loop-e-deformed} ]{\includegraphics[width=.3\textwidth]{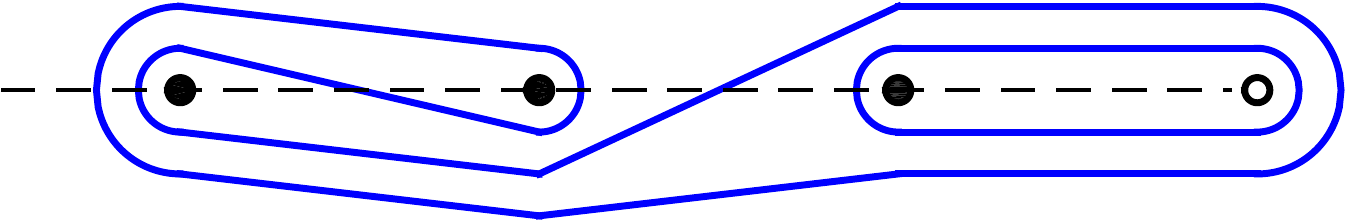}}
  \caption{Top: The loop \(\ell_{E}\) for the \(3\)-punctured disk, used in
    the iterative computation of topological
    entropy~\eqref{eq:braid-entropy}. The three main punctures are black,
    while the white puncture serves as the ``basepoint'' for the loop.
    Bottom: \(\ell_{E}\) after deformation by the braid
    \(\sigma_{1}\sigma_{1}\), which exchanges the first pair of punctures
    twice, clockwise.}
  \label{fig:generator-loop}
\end{figure}

Thurston~\cite{Fathi1979} showed that \(h(\mathbf{b})\) can be computed from
the growth rate of a loop \(\ell_{E}\) under repeated action of the braid
\(\mathbf{b}\):
\begin{equation}
  \label{eq:braid-entropy}
  h(\mathbf{b})
  = \lim_{k \to \infty} \frac{1}{k}
  \log \frac{\abs{\mathbf{b}^{k} \ell_{E}}}{\abs{\ell_{E}}}.
\end{equation}
Here we take the loop \(\ell_{E}\) to be a generating set for the fundamental
group of the \(n\)-punctured disk \(\mathbb{D}\).~\cite{Moussafir2006} It can
be represented in Dynnikov coordinates by adding an extra ``basepoint''
puncture that does not participate in the braid \(\mathbf{b}\)
(\autoref{fig:generator-loop}).  The length \(\abs{\cdot}\) is the number of
intersections of the loop with the horizontal axis to the left of the
basepoint (dashed line in \autoref{fig:generator-loop}).

This iterative calculation may however be an inappropriate measure of
complexity of trajectories. First, unless the trajectories correspond to
periodic orbits, there is no justification for repeated application of the
braid, since in that case the physical braid does not represent a
homeomorphism of a fixed punctured surface.\footnote{See Section~5 of the
  \texttt{braidlab} guide~\onlinecite{Thiffeault2014v3} for a consequence
  of removing strings from braids of non-periodic trajectories.}
Second, flows that stretch material sub-exponentially will always be assigned
\(h(\mathbf{b}) = 0\), which may be too crude of a measure in practice, e.g.,
for estimating time-scales for transport.

Dynnikov and Wiest\cite{Dynnikov2007} used a non-iterated version
of~\eqref{eq:braid-entropy} to study the connection between algebraic and
geometric properties of braids. We adapt their definition here to define the
Finite-Time Braiding Exponent as a measure of complexity of trajectories.
\begin{definition*}[Finite-Time Braiding Exponent]
  Let \(\mathbf{b}\) be a braid corresponding to \(n\) trajectories over a
  time interval of length \(T\).  The Finite-Time Braiding Exponent is given
  by
  \begin{equation}
    \label{eq:ftbe}
    \ftbe(\mathbf{b})
    = \frac{1}{T}\,\log \frac{\abs{\mathbf{b} \ell_{E}}}{{\abs{\ell_{E}}}}
  \end{equation}
  where \(\ell_{E}\) is the loop representing a generating set of the
  non-oriented fundamental group on \(n\)-punctured disk.
\end{definition*}
A similar definition of exponential deformation rate appeared in
ref.~\onlinecite{Thiffeault2010}, which uses an arbitrary loop \(\ell\)
instead of \(\ell_{E}\).

\section{Constructing braids and computing FTBEs}
\label{sec:characterization}

Our ultimate goal is to use FTBEs as an approximation to topological entropy
of a planar dynamical system.  In this section we discuss the issues involved
in constructing braids from data, introduce our model system, and examine the
sensitivity of braids and FTBEs to various model parameters.

\subsection{Practical considerations}

To compute FTBEs, we must first construct braids.  To do this, we record a
finite number of concurrent, non-intersecting trajectories of the
dynamics. Such a data set is determined by three parameters:
\begin{compactenum}[1)]
\item the number \(n\) of trajectories,
\item the initial conditions of trajectories, and
\item the time interval \([t_{0},t_{0}+T]\) during which trajectories are
  recorded.
\end{compactenum}
The MATLAB toolbox \texttt{braidlab}~\cite{Thiffeault2014v3} converts
trajectories into braids and then computes FTBEs. \texttt{braidlab} is an
open-source library that implements a wide set of operations on braids, loops,
and related structures. In particular, braids generated from data as discussed
here are stored as \texttt{databraid} data structures, which record both
crossings of strands as generators and the times at which crossings occurred.

The computational time to convert \(n\) trajectories to a braid scales
quadratically with \(n\), and linearly with both the length of trajectories
and the total number of crossings between strands. The algorithm depends on
two additional parameters:
\begin{compactenum}[1)]
  \setcounter{enumi}{3}
\item the angle \(\alpha\) of the projection axis used to assign order to
  strands; and
\item the time step \(\tau\) at which trajectories are sampled, assuming
  uniform sampling.
\end{compactenum}
To accurately convert discretized trajectories to a braid, \texttt{braidlab}
requires that strands cross only between sampled times, and that any two
strands cross at most once per time step.  If strands cross exactly at the
moment of sampling, a braid generator cannot be unambiguously assigned. Such
degeneracies may be resolved by varying the angle
\(\alpha\)\cite{Thiffeault2014v3}.

\begin{figure}[h!]
  \centering
  \subfigure[\ Identity (trivial) braid]{\includegraphics[width=.5\columnwidth]{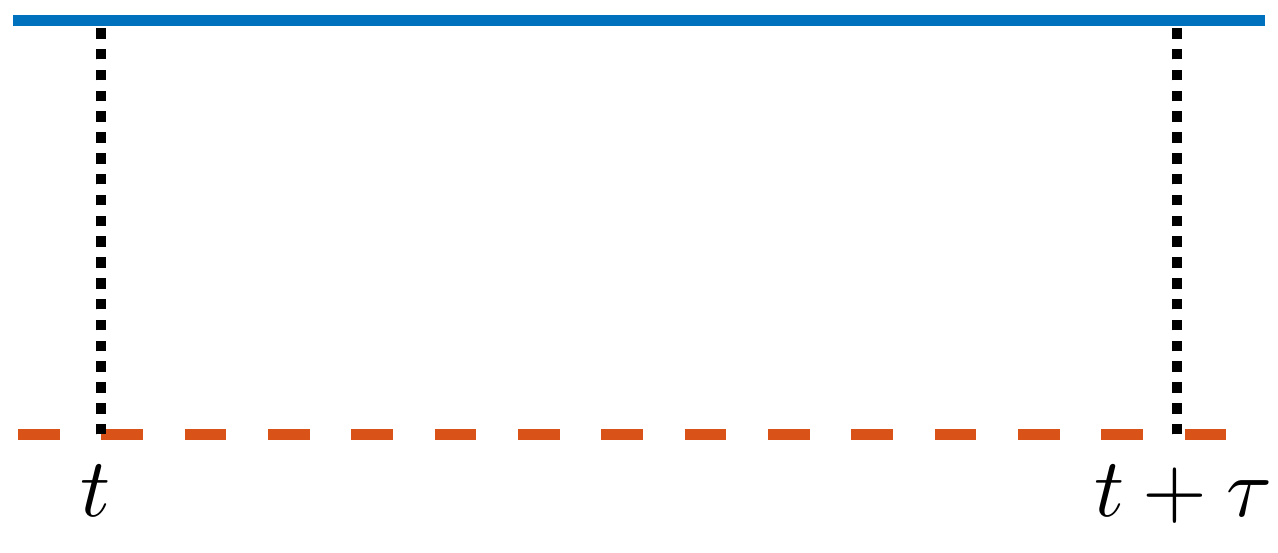}}
   \subfigure[\ \(\sigma_{i}\sigma_{i}^{-1}\) braid]{\includegraphics[width=.5\columnwidth]{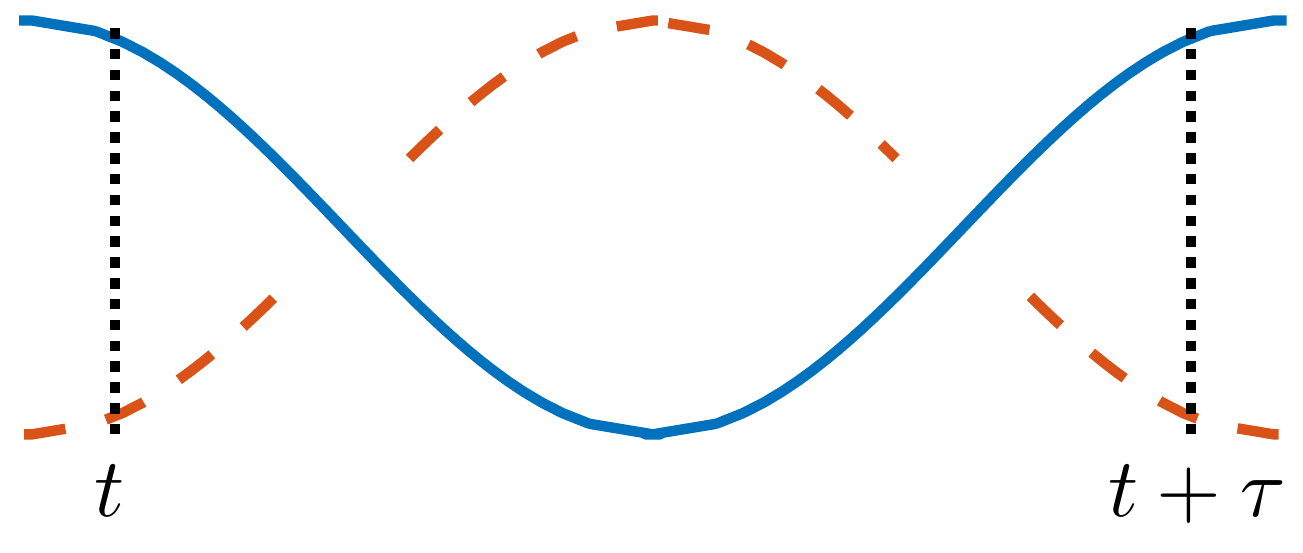}}
   \subfigure[\ \(\sigma_{i}\sigma_{i}\) braid]{\includegraphics[width=.5\columnwidth]{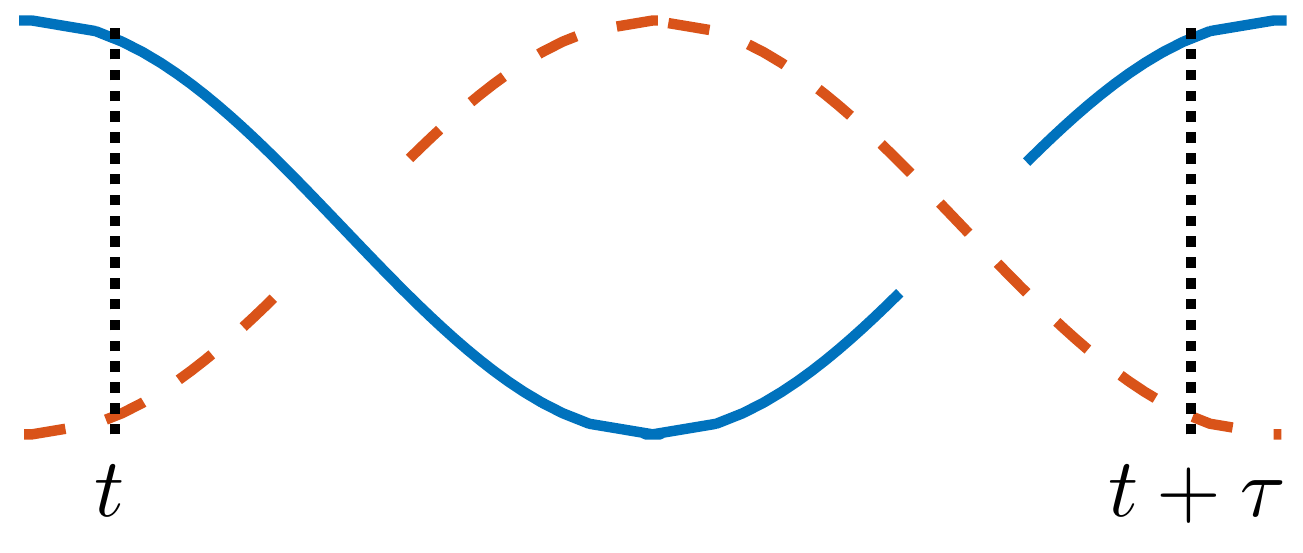}}
  \caption{Three physical braids with same orders of strands \(i\), \(i+1\) at sampling times \(t\) and \(t+\tau\), but different sequences of generators. Using time step \(\tau/2\) would resolve the crossings correctly.}\label{fig:crossings}
\end{figure}

When two strands cross more than once in a single time step, other ambiguities
can arise. For example, \autoref{fig:crossings} shows three different physical
braids with same strand positions at the sampling time steps.  The ambiguity
can clearly be resolved by reducing \(\tau\).

The sections that follow demonstrate the dependence of FTBEs on each of the
five parameters listed above: number of strands \(n\), initial
conditions, duration of time interval \(T\), projection
angle \(\alpha\), and time step \(\tau\). The input trajectories are
generated by a mixing dynamical system, the \arefmodel{}.

\subsection{The \arefmodel{}}
\label{sec:aref-blinking-vortex}

\begin{figure}[htb]
  \centering
  \includegraphics{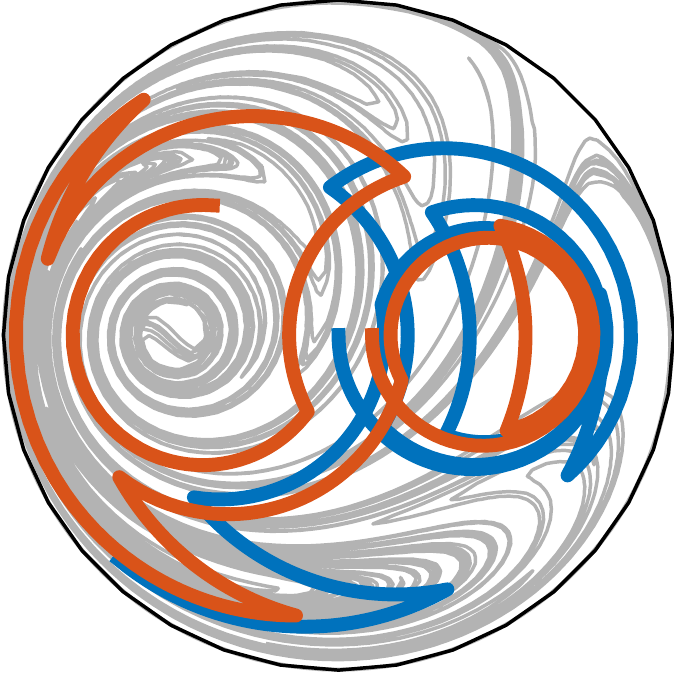}
  \caption{Material advection patterns and two particle trajectories of the \arefmodel{} in the mixing regime.}\label{fig:aref-flow}
\end{figure}

The \arefmodel{}\cite{Aref1984} is an idealization of a device that stirs
fluid using two counter-rotating vortices in a circular domain.  The vortices
are otherwise identical, and are positioned along the diameter of the unit
circle at distances \(\pm b\) from its center. They are turned on alternately
(``blinked'') during half of the period \(T_{P}\) of the protocol, resulting
in continuous, piecewise-differentiable trajectories of fluid parcels, shown
in \autoref{fig:aref-flow}. After fixing the geometry, and period of the
protocol to \(T_{P}=1\), the only free parameter is the non-dimensional
circulation \(\mu\). We restrict our analysis to \(\mu \in [3,9]\) for which
numerical experiments indicate that the flow is mixing in the entire domain.\cite{Aref1984}

We quantify the complexity of the flow by estimating topological entropy \(h\) using direct advection of material curves~\cite{Newhouse1993}. Numerically, we seed initial conditions along a randomly selected straight material line and advect those points forward. When distances between neighboring points grow too large, we linearly interpolate additional points between them to ensure that details of bends in the material curve are well represented.

\begin{figure}[h!]
  \centering
  \includegraphics{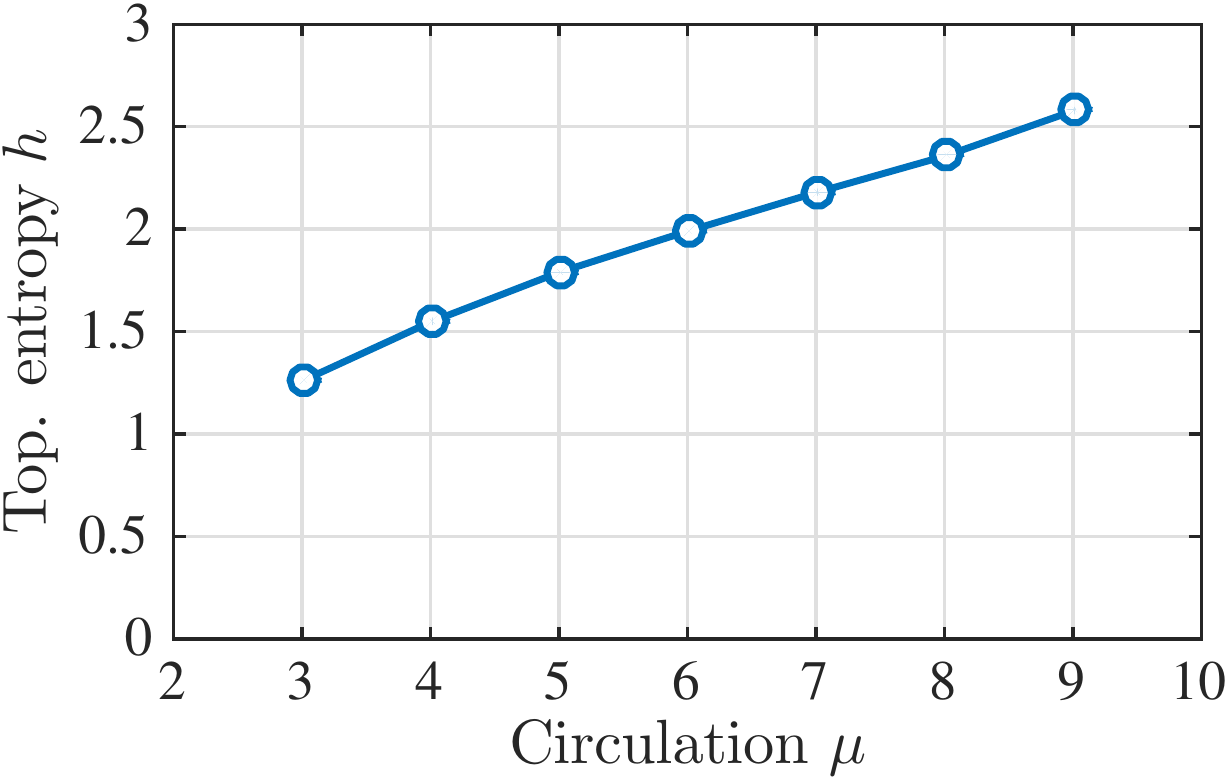}
  \caption{Topological entropy of counter-rotating \arefmodel{}}\label{fig:topentropy}
\end{figure}

\autoref{fig:topentropy} shows the estimated values of topological entropy \(h\) that will be used as reference for our later calculations of FTBEs.  Topological entropy \(h\) is approximately linear in circulation \(\mu\), which determines the dominant time scale of the flow.

\subsection{Robustness of braid construction}
\label{sec:param-braids}

In this section we investigate numerical issues involved in braid construction,
which could lead to poorly-defined braids if not treated properly.  The two
main issues are the choice of time step (sampling rate,
\autoref{sec:time-step}) and the choice of projection angle
(\autoref{sec:projection-angle}).

\subsubsection{Time step}
\label{sec:time-step}

Let us investigate how the choice of the sampling interval \(\tau\) affects
accuracy of constructed braids and their FTBEs. We will use the word length of
braids as a proxy for the accuracy of constructed braids, as reducing \(\tau\)
increases the number of crossings in the braid.

\begin{figure}[h!]
  \centering
  \includegraphics{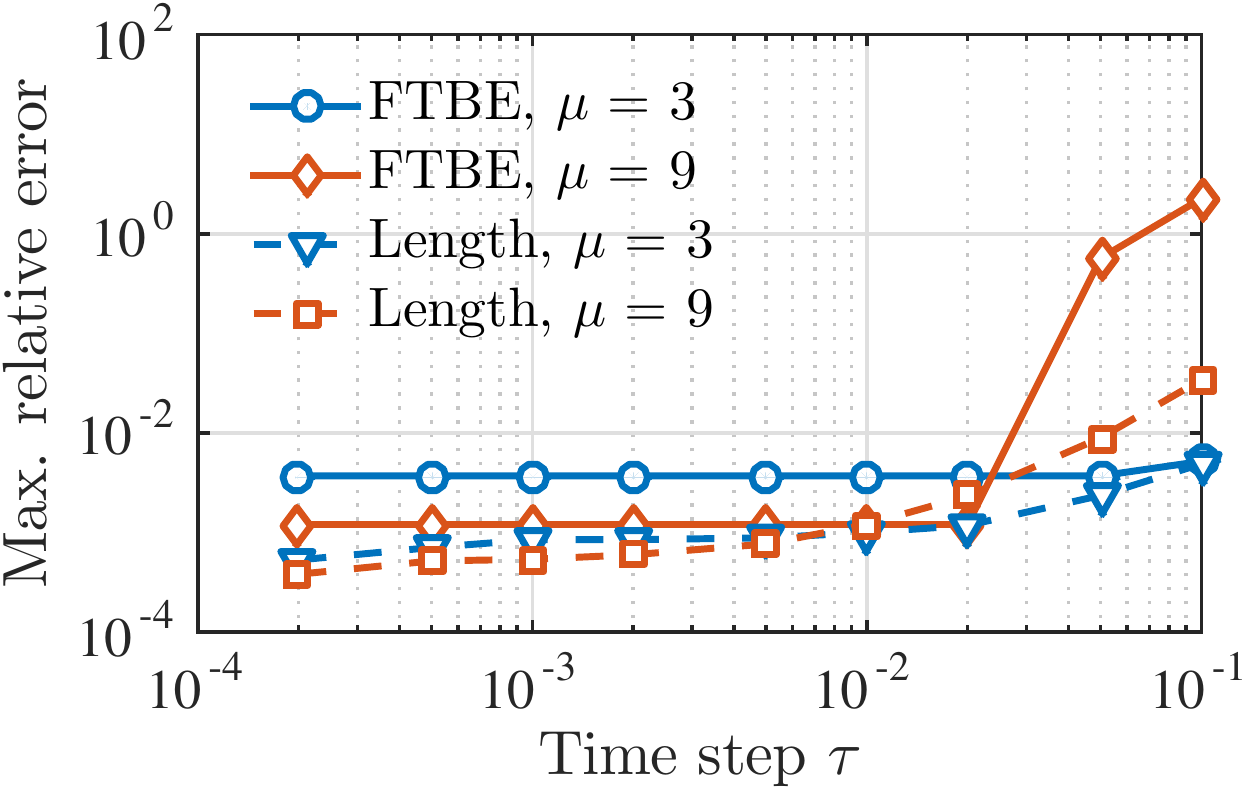}
  \caption{Effect of time step \(\tau\) on relative error in FTBE and braid
    word length~\eqref{eq:relative-errors-timestep} with respect to the
    reference time step \(\tau_{\ast} = 10^{-4}\). Graph shows the maximum
    error among \(100\) braids of 50 uniformly initialized
    strands.}\label{fig:dependence-on-timestep}
\end{figure}

Vortices in \arefmodel{} are modeled by singularities which we slightly
regularize to avoid infinite velocities. After regularization, time step
\(\tau_{\ast} = 10^{-4}\) is sufficient to accurately sample trajectories of
the flow in the studied regime of circulations. For each set of initial
conditions we compute a braid \(b_{\ast}\) with time step \(\tau_{\ast}\), which serves
as a reference, and additional braids \(b_{\tau}\) with larger values of time
step \(\tau\). Figure~\ref{fig:dependence-on-timestep} shows how the relative errors
\begin{equation}
  \label{eq:relative-errors-timestep}
  \abs{1 - \frac{\ftbe(\mathbf{b}_{\tau})}{\ftbe(\mathbf{b}_{\ast})}}, \text{ and }
  \abs{1 - \frac{L(\mathbf{b}_{\tau})}{L(\mathbf{b}_{\ast})}},
\end{equation}
depend on time step \(\tau\). To reduce the effect of spatial
variations, we plot the largest relative errors with among different uniformly-initialized braids.

Figure~\ref{fig:dependence-on-timestep} suggests that FTBEs are more resilient
to large time steps than word length. This implies that some parts of the
braid do not contribute significantly to deformation of loops. For example, a
braid of four generators \(\sigma_{i} \sigma_{i}^{-1} \sigma_{i}
\sigma_{i}^{-1}\) introduces no net deformation as each inverse generator
reverses the action of the previous one. Such sequences can be generated by
pairs of strands that are neighbors along the projection line, but are
otherwise located far from each other in the domain.

As expected, the effect of large time steps is more pronounced for flows with
larger circulations \(\mu\), as particles move farther within each time
step.  We use \(\tau = 10^{-2}\) for the rest of the paper, which is
appropriate for our range of \(\mu\).

\subsubsection{Angle of the projection axis}
\label{sec:projection-angle}

The angle \(\alpha\) determines the projection line used to assign order
indices to strands of a physical braid. As mentioned at the opening of
Section~\ref{sec:characterization}, \(\alpha\) may need to be adjusted to
ensure that trajectories can be converted to a braid. We show that even though
the braid itself changes depending on \(\alpha\), the effect on FTBEs is
minimal.

When a physical braid consists of periodic trajectories, changing \(\alpha\)
results in conjugation of the corresponding braid \(\mathbf{b}\) by another
braid \(\mathbf{c}_{\alpha}\). In other words, dependence on angle \(\alpha\)
enters into the braid as \(\mathbf{c}_{\alpha}^{-1} \mathbf{b}
\mathbf{c}_{\alpha}\). However, when strands in the physical braid are not
periodic, changing \(\alpha\) is not guaranteed to affect the braid through
conjugation only~\cite{Thiffeault2014v3}. Nevertheless, since changing
\(\alpha\) only rotates the coordinate system, the FTBEs should not
significantly depend on~\(\alpha\).

\begin{figure}[h!]
  \centering
  \subfigure[\ Mean FTBE]{ \includegraphics{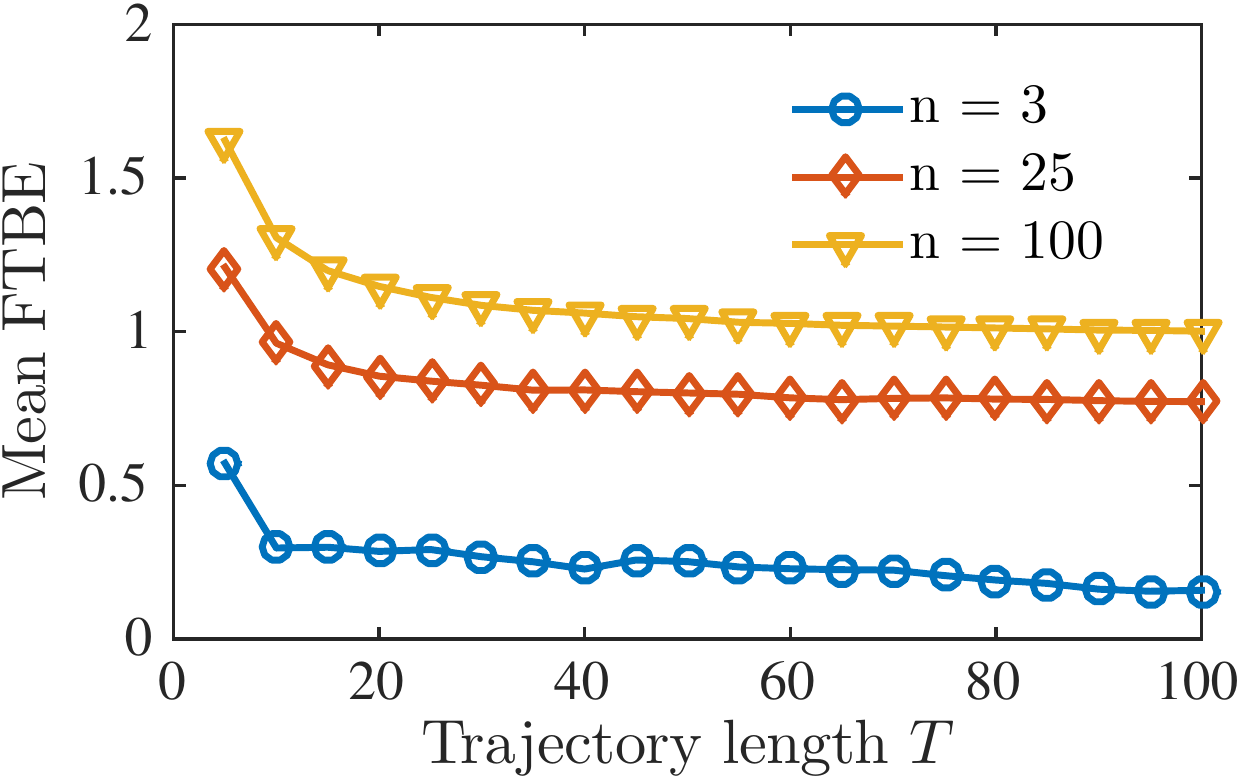} }
  \subfigure[\ Relative Standard Deviation of FTBE]{\includegraphics{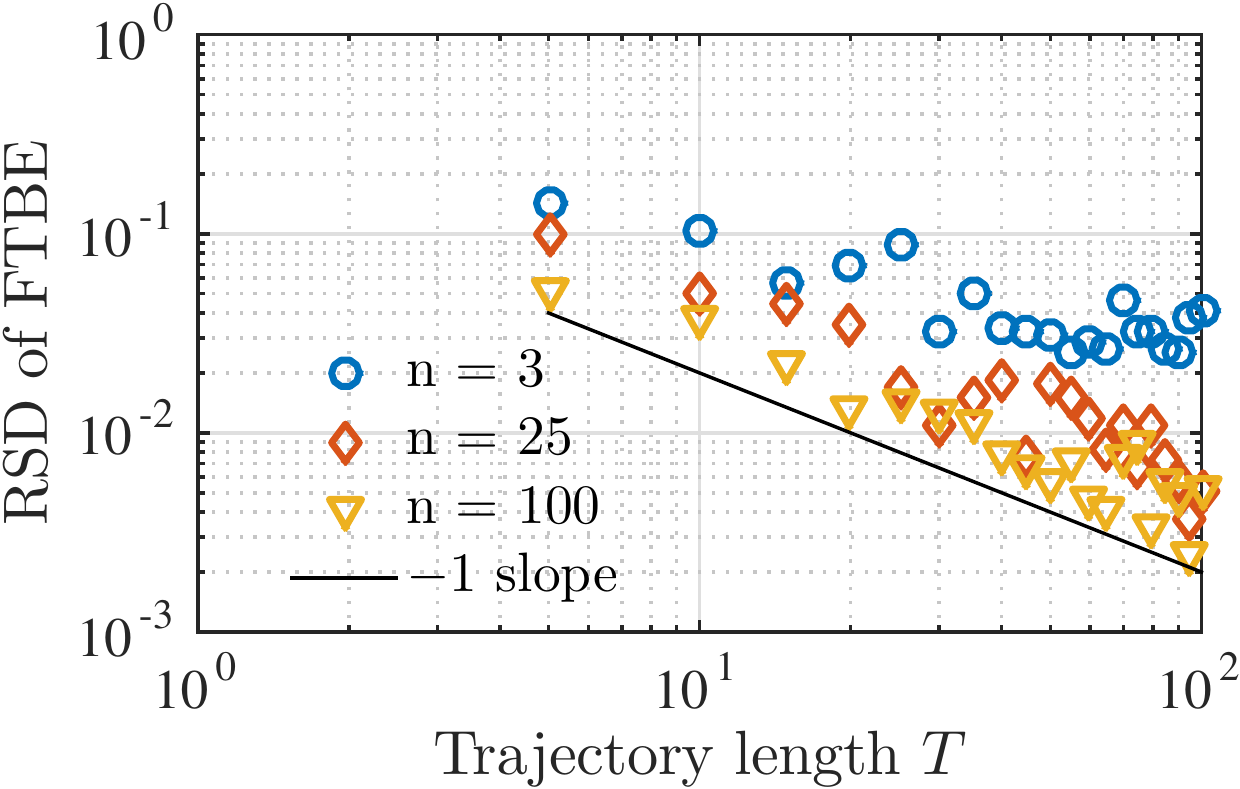}}
  \subfigure[\ Mean braid word length]{\includegraphics{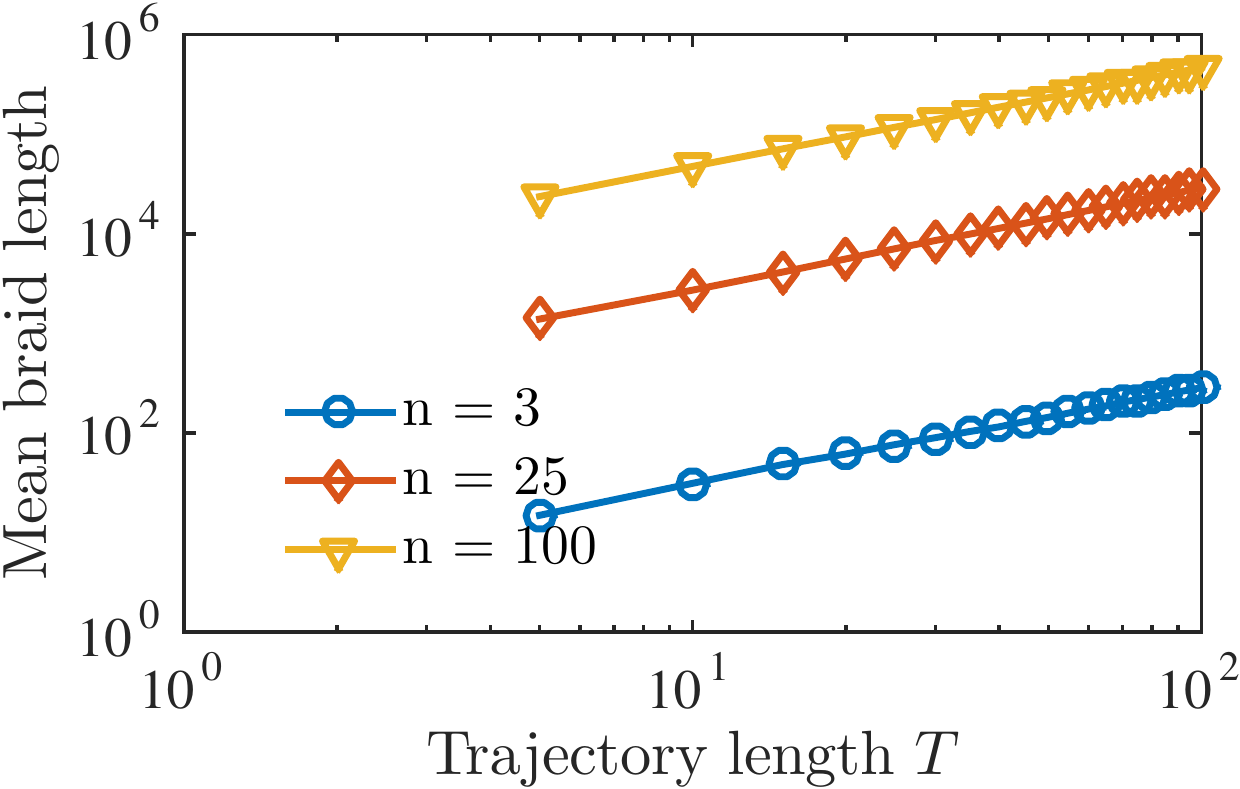}}
  \subfigure[\ Relative Standard Deviation of braid word length]{\includegraphics{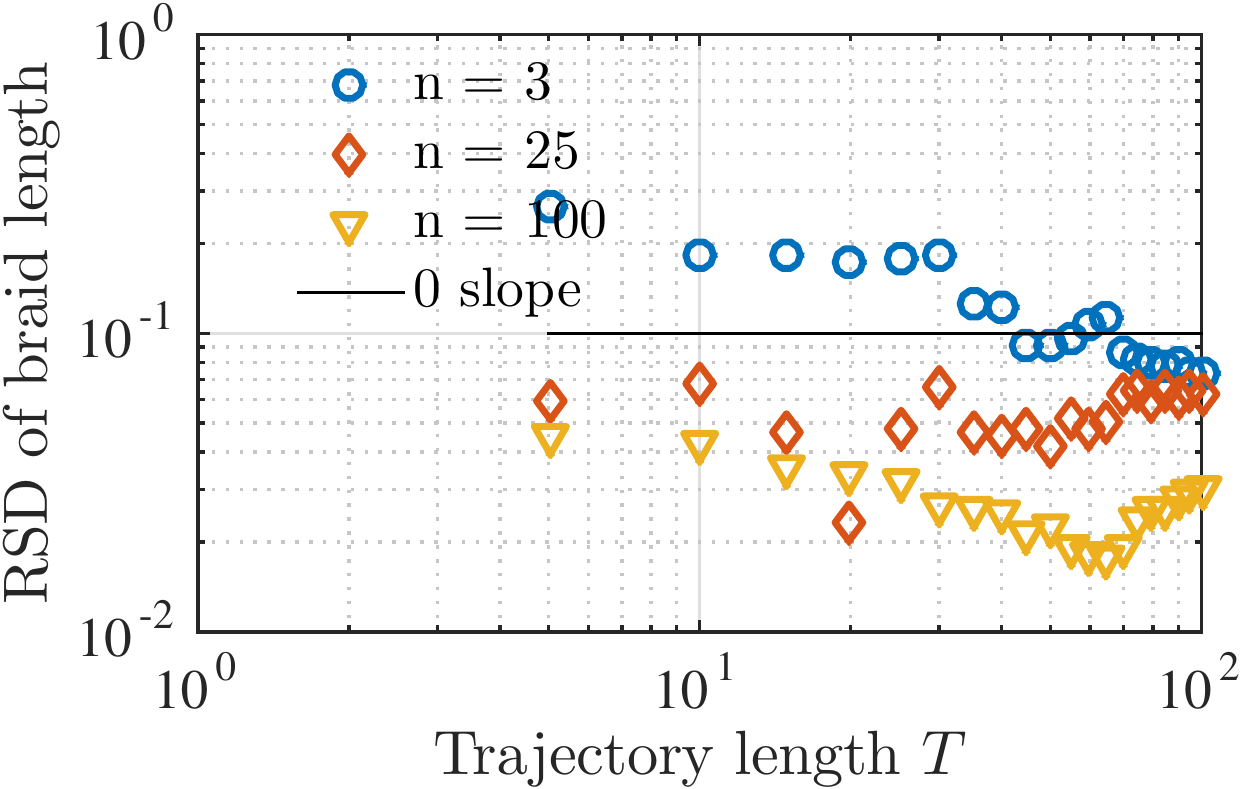}}
  \caption{Effect of projection angle \(\alpha\) on FTBEs and braid word length of a single physical braid. Statistics are computed with respect to uniform distribution of~\(100\) angles \(\alpha \in [0,\pi)\). Circulation is \(\mu = 5\).}\label{fig:angle-dependence}
\end{figure}

\autoref{fig:angle-dependence} shows how length \(T\) of trajectories
influences statistics of FTBEs and braid word length with respect to
projection angle \(\alpha\). It is clear that the change of projection angle
minimally affects the value of FTBEs for longer braids. Over the entire range
of times \(T\) simulated, the Relative Standard Deviation\footnotemark[4]
(RSD) of FTBEs decays inverse-proportionally with length of the physical
braids, while the standard deviation of length remains approximately constant,
indicating that FTBEs are more robust than braid length with respect to change
of \(\alpha\). Based on this, we set \(\alpha=0\) in the remainder of this
paper.\footnotetext[4]{Relative
  Standard Deviation (RSD) is the standard deviation divided by the mean.}

\subsection{Parameters for braid construction}
\label{sec:param-data-collection}

In \autoref{sec:param-braids} we addressed numerical issues that can lead to
deviations from the idealized mathematical representation of braids.  In this
section we turn to parameters that affect the nature of the braid we
obtain.  These are the choice of initial conditions, the length (duration)
of trajectories (\autoref{sec:init-cond-length}), and the number of
trajectories (\autoref{sec:dependence-on-nstrands}).

\subsubsection{Initial conditions and length of trajectories}
\label{sec:init-cond-length}

FTBEs can be interpreted as measures of \emph{average deformation} of a loop
over finite segments of physical braids. Averaged measurements in chaotic
systems quickly stabilize to constant values independent of precise initial
conditions, which is formally asserted by various flavors of ergodic theorems.
Since numerical evidence indicates that the \arefmodel{} is mixing for \(\mu
\in [3,9]\), the FTBEs should not depend on initial conditions for long-enough
trajectories.

\begin{figure}[h!]
  \centering
  \subfigure[\ Mean FTBE]{\includegraphics{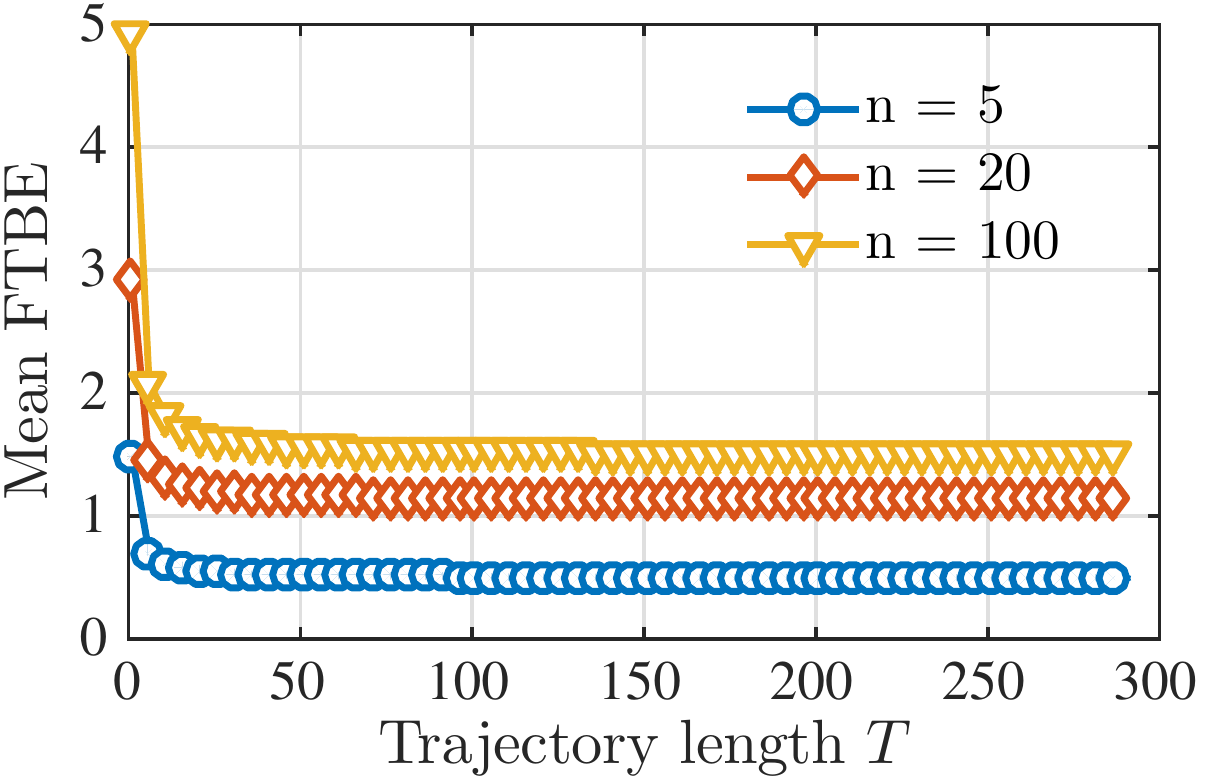}}
  \subfigure[\ Relative Standard Deviation of FTBE]{\includegraphics{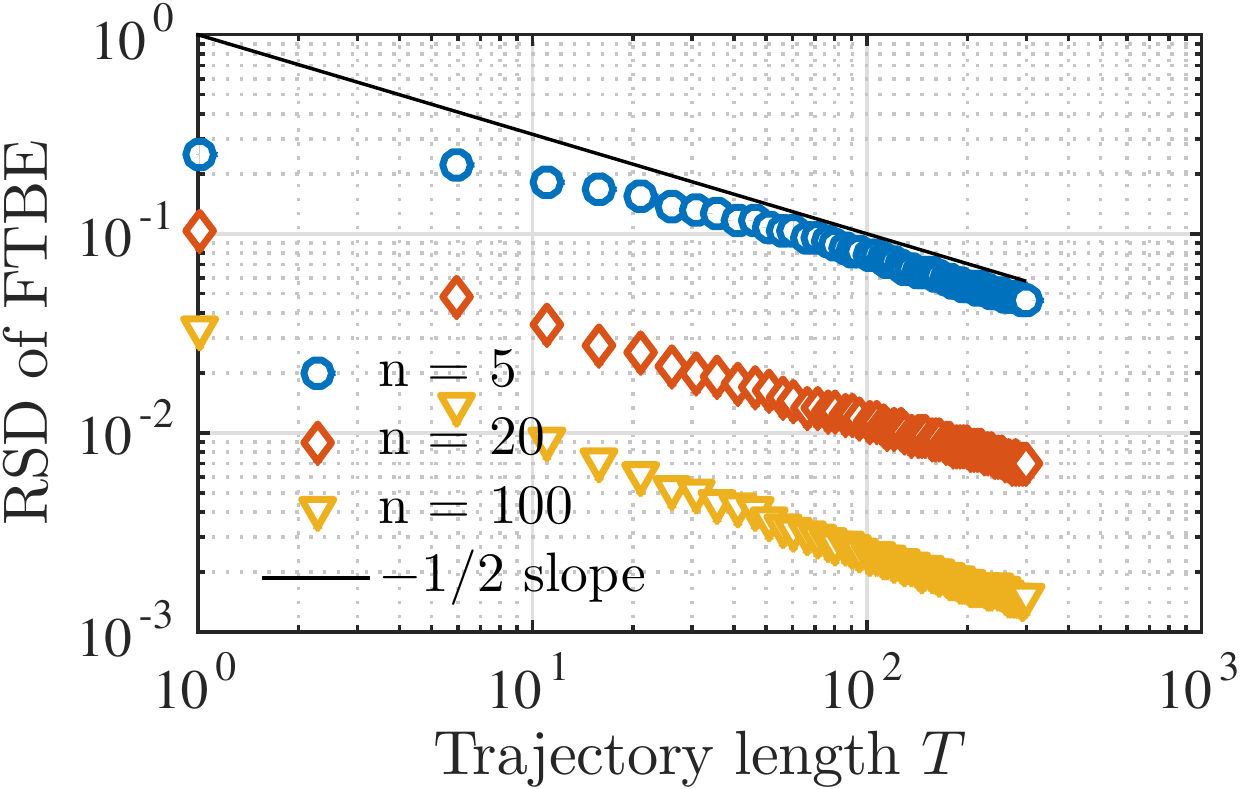}}
  \caption{Effect of trajectory length \(T\) on FTBE for an ensemble of physical braids. Statistics are computed with respect to uniform initialization of 100 physical braids. The circulation is \(\mu = 5\). }\label{fig:time-dependence}
\end{figure}

\autoref{fig:time-dependence} confirms the expected ergodic behavior. After the rapid initial transient, the spatial mean of FTBEs settles to a constant value, while spatial fluctuations decay following a power law. Since the length of the transient does not depend on number \(n\) of strands, we set the length of trajectory to \(T = 100\) in future simulations.

The slope \(-1/2\) of decay of fluctuations is commonly associated with the
Central Limit Theorem for sums of random variables. Variants of the Central
Limit Theorem apply to the decay in fluctuations of Finite-Time Lyapunov
Exponents in mixing flows~\cite{Hennion1997,Young1998}. While we provide no
rigorous proof that the same applies to FTBEs, it is not surprising to see
the same behavior since FTBEs and FTLEs are closely-related quantities.

\subsubsection{Number of strands}
\label{sec:dependence-on-nstrands}

It has been observed that using more trajectories increases the complexity of
the braid~\cite{Finn2007, Thiffeault2005, Tumasz2012a}. To study how FTBEs
depend on the number of strands \(n\), we fix the length of trajectories to
\(T = 100\) and at each value \(n\) compute mean and standard deviation\footnotemark[4] of the
FTBE with respect to initial conditions.

\begin{figure}[h!]
  \centering
  \subfigure[\ Mean FTBE. Horizontal lines indicate topological entropies \(h\).]{\includegraphics{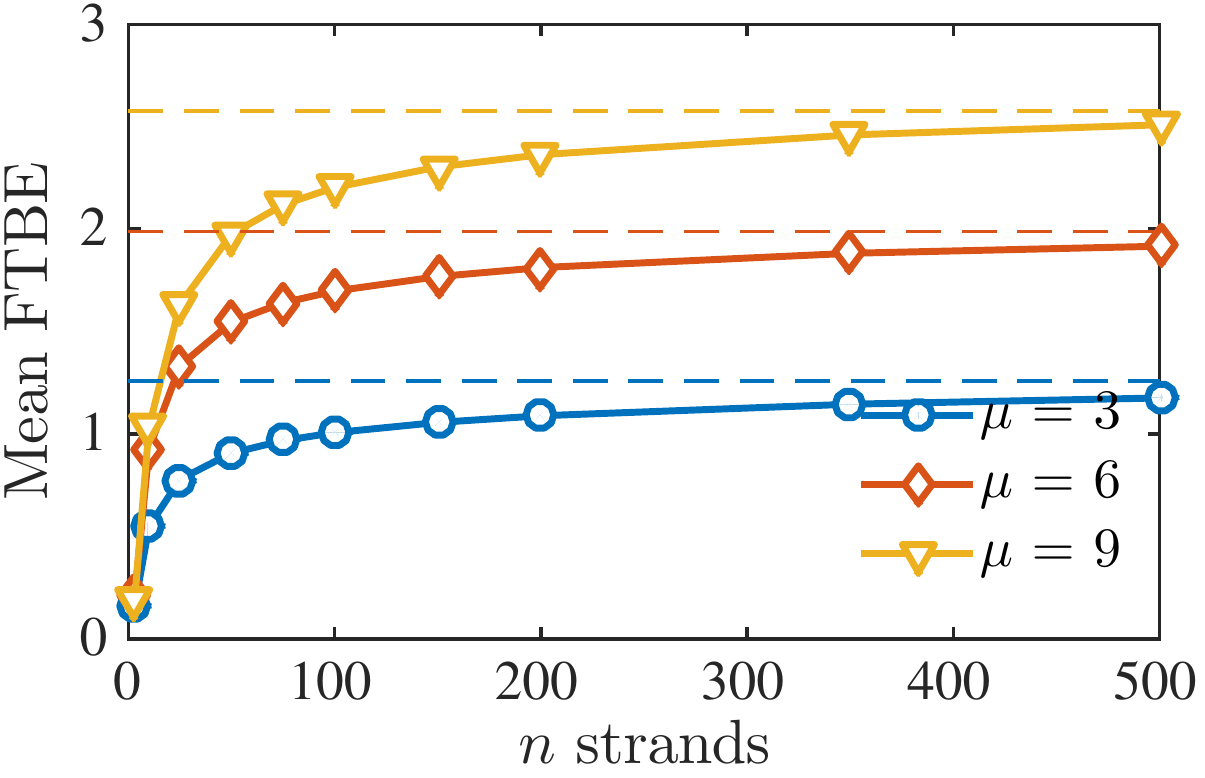}}
  \subfigure[\ Relative Standard Deviation of FTBE\label{fig:strands-dependence-ftbe-std}]{\includegraphics{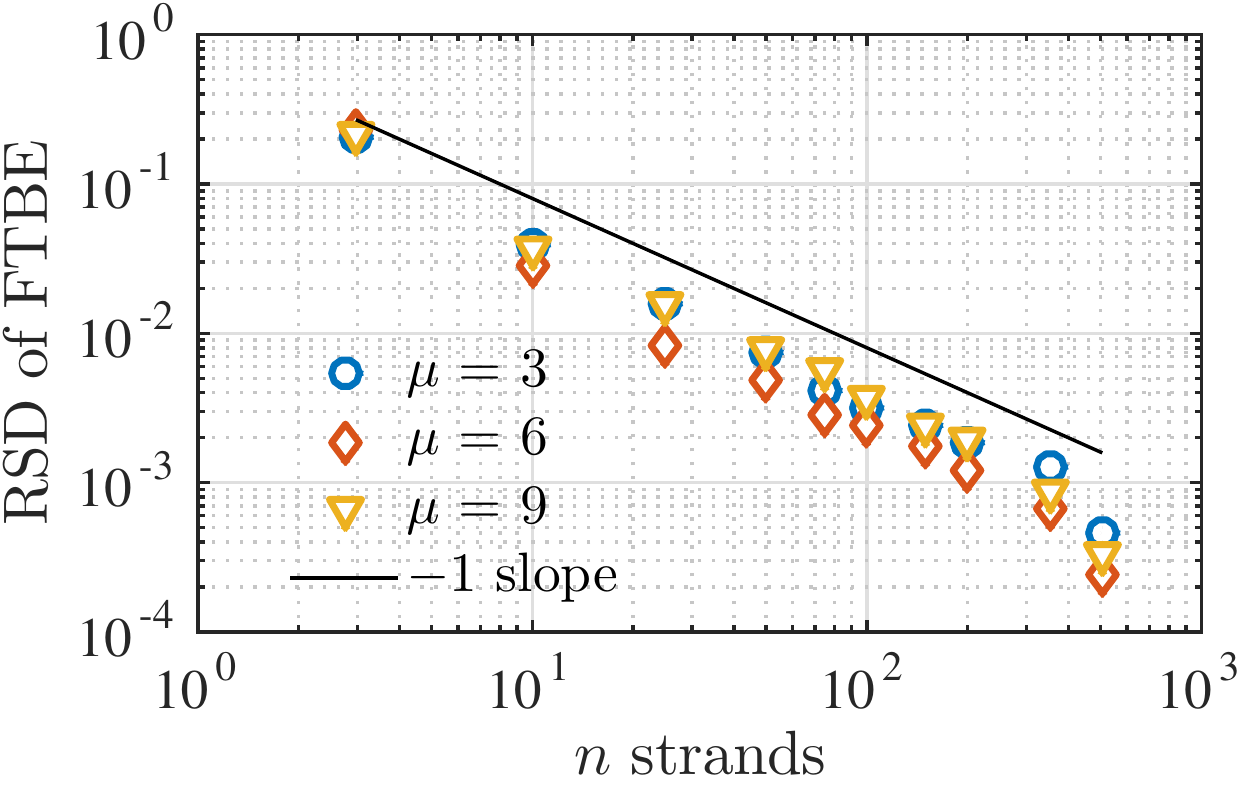}}
  \subfigure[\ Mean braid length]{\includegraphics{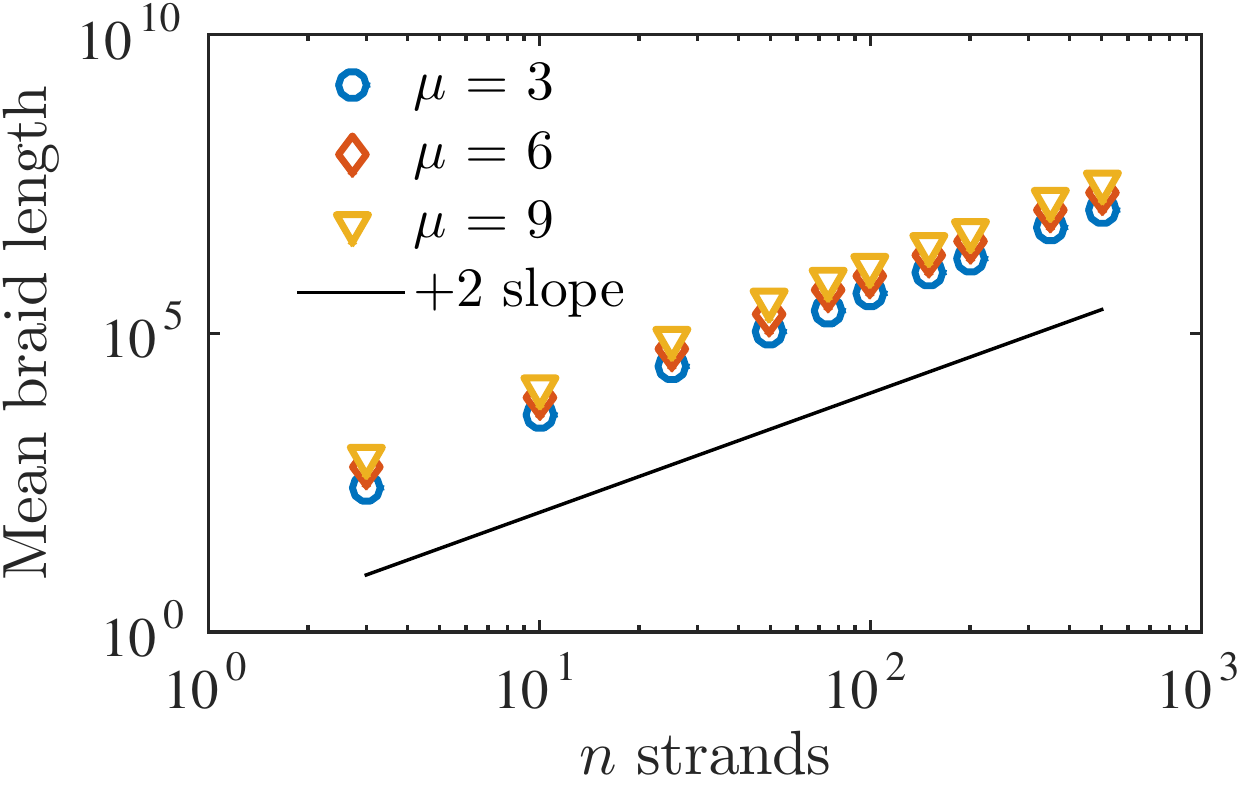}}
  \subfigure[\ Relative Standard Deviation of braid length]{\includegraphics{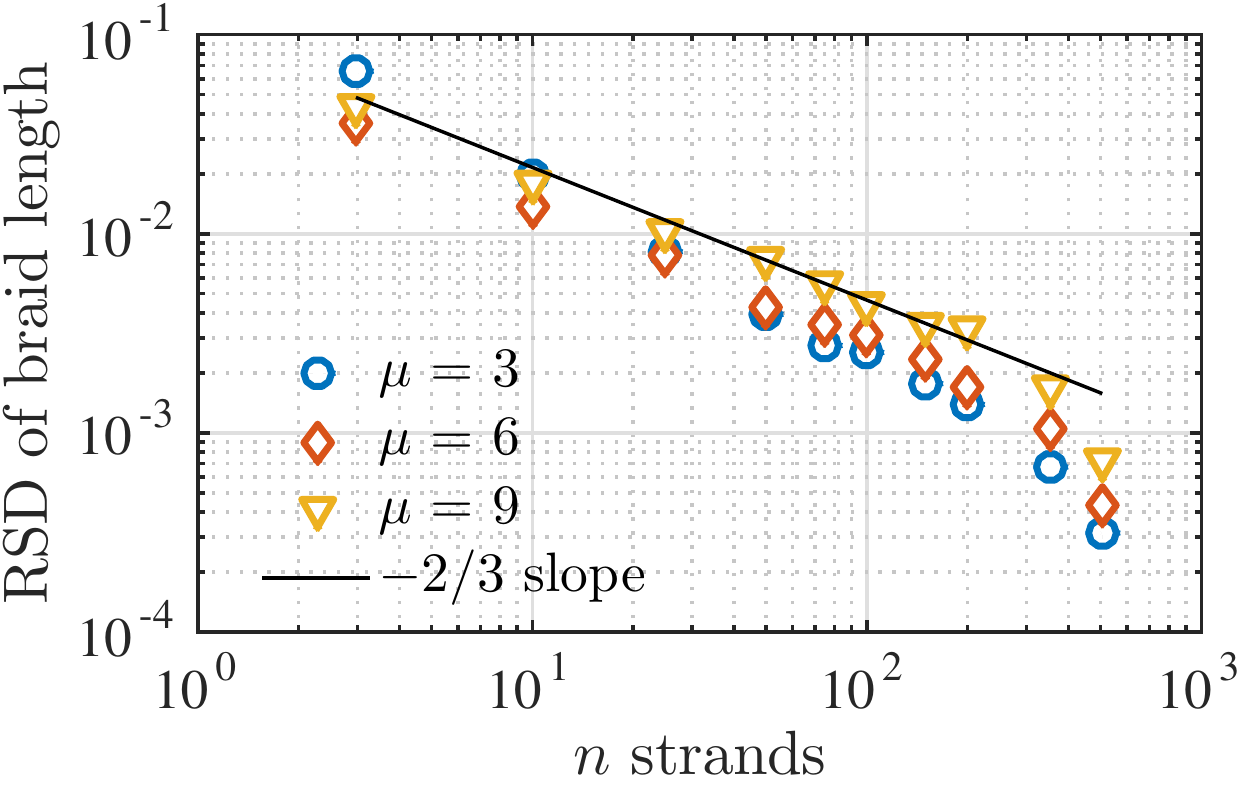}}
  \caption{Influence of number \(n\) of strands on FTBE and length of an ensemble of physical braids. Statistics are computed with respect to uniform initialization of 100 physical braids.}\label{fig:strands-dependence}
\end{figure}

\autoref{fig:strands-dependence} shows that mean \(\ftbe\) and mean braid word
length grow in a different manner with addition of strands. The mean \(\ftbe\)
appears to limit to topological entropy \(h\); on the other hand, the number of
crossings \(L\) is unbounded, growing proportionally to the number of pairs
of strands \(n^{2}\). The standard deviations of both FTBE and word length
decay according to a power law, although with slightly different slopes.

To generate \autoref{fig:strands-dependence} we computed approximately \(S
\times \sum_{\mathrm{points}} n \approx 150,000\) trajectories of the
\arefmodel{}. While such a large number of trajectories can be computed for
numerical flows, the number of trajectories available experimentally is often
much smaller, especially in systems featuring human crowds, flocks of birds,
or oceanic floats. We can, however, generate collections of braids by re-using
data from a much smaller set of trajectories.

If we have access to \(m\) measured trajectories, we can form a single
\(m\)-stranded braid, but also select subsets of \(n < m\) trajectories to form
smaller \(n\)-\emph{\subbraid{}s}.  There are \({m \choose n}\) different
\(n\)-stranded \subbraid{}s, which is a number that grows fast as \(n \to m/2\) from
above or below. Therefore, it is easy to obtain large collections of
\(n\)-braids by selectively including strands of a bigger set of \(m\)
trajectories.

\begin{figure}[h!]
  \centering
  \subfigure[\ Mean FTBE]{\includegraphics{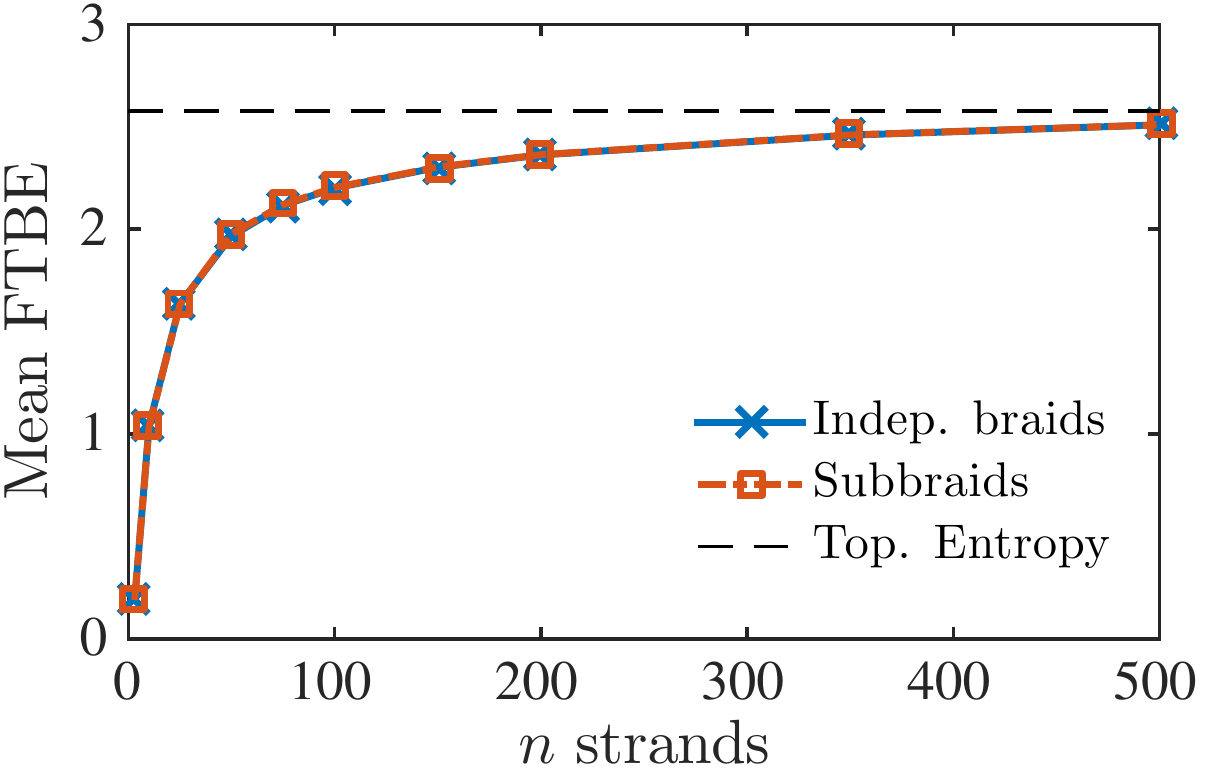}}
  \subfigure[\ Relative Standard Deviation of FTBE\label{fig:subbraid-comp-ftbe-std}]{\includegraphics{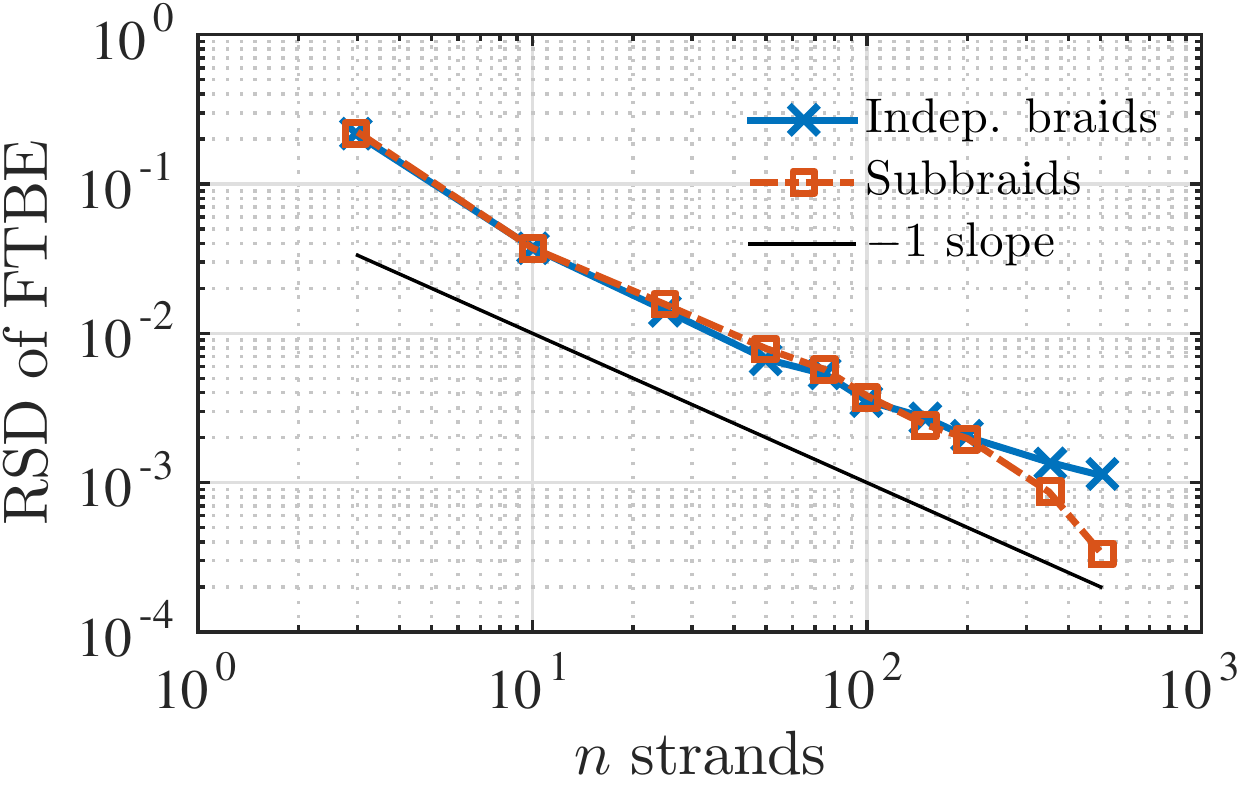}}
  \subfigure[\ Distribution of FTBE at \(n = 500\), value with largest difference in RSD between two sets.]{\includegraphics{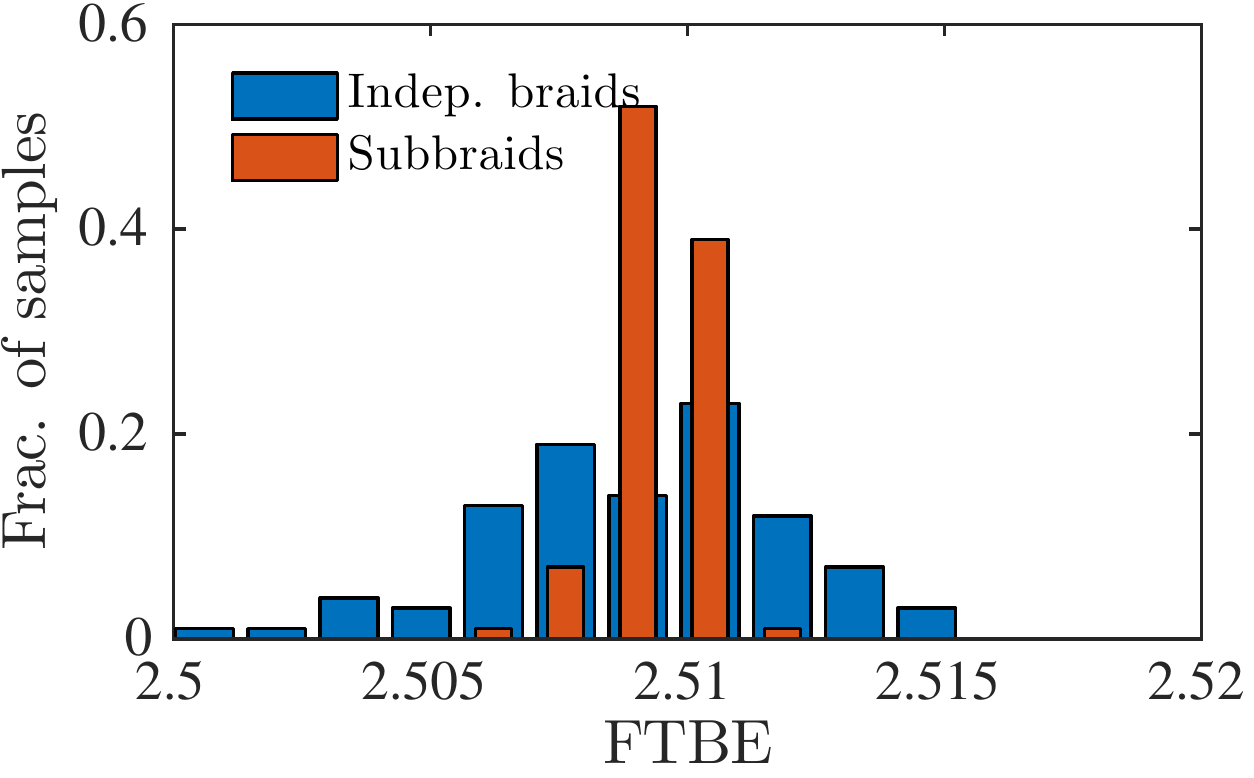}}
  \caption{Comparison of FTBE for braids sampled independently and braids
    formed as \subbraid{}s of \(m\) strands chosen from a single \(m=550\)
    braid. Sample size at every parameter value is \(S = 100\), circulation
    \(\mu = 9\).}\label{fig:subbraid-comp}
\end{figure}

Despite the practical appeal, it is not clear that statistics of FTBEs and
braid lengths will be similar if calculated using either independent braids or
\subbraid{}s of a larger braid; \subbraid{}s are not mutually independent, as
they re-use the same set of trajectories from the original data set. To
compare statistics, we generate \subbraid{}s from a single set of \(m=550\)
trajectories for the same values of \(n\) used in the calculations
\autoref{fig:strands-dependence}, which used independent
sampling. Figure~\ref{fig:subbraid-comp} overlays the results of two
calculations and indicates that statistics of FTBEs over \subbraid{}s match
almost perfectly with statistics over independently sampled braids.

An additional practical benefit of using \subbraid{}s is the speed of
generating them. The time to extract one \(n\)-\subbraid{} from a full
\(m\)-braid scales with the braid word length of the full braid. This
procedure is generally much faster than forming an independent \(n\)-braid,
even if the time to simulate new trajectories is negligible. In our example
statistics over \subbraid{}s were computed \(5\) to \(10\) times faster than
over independent braids.

\section{Extrapolating topological entropy from FTBEs}
\label{sec:extrapolation}

\begin{figure}[h!]
  \centering
  \includegraphics{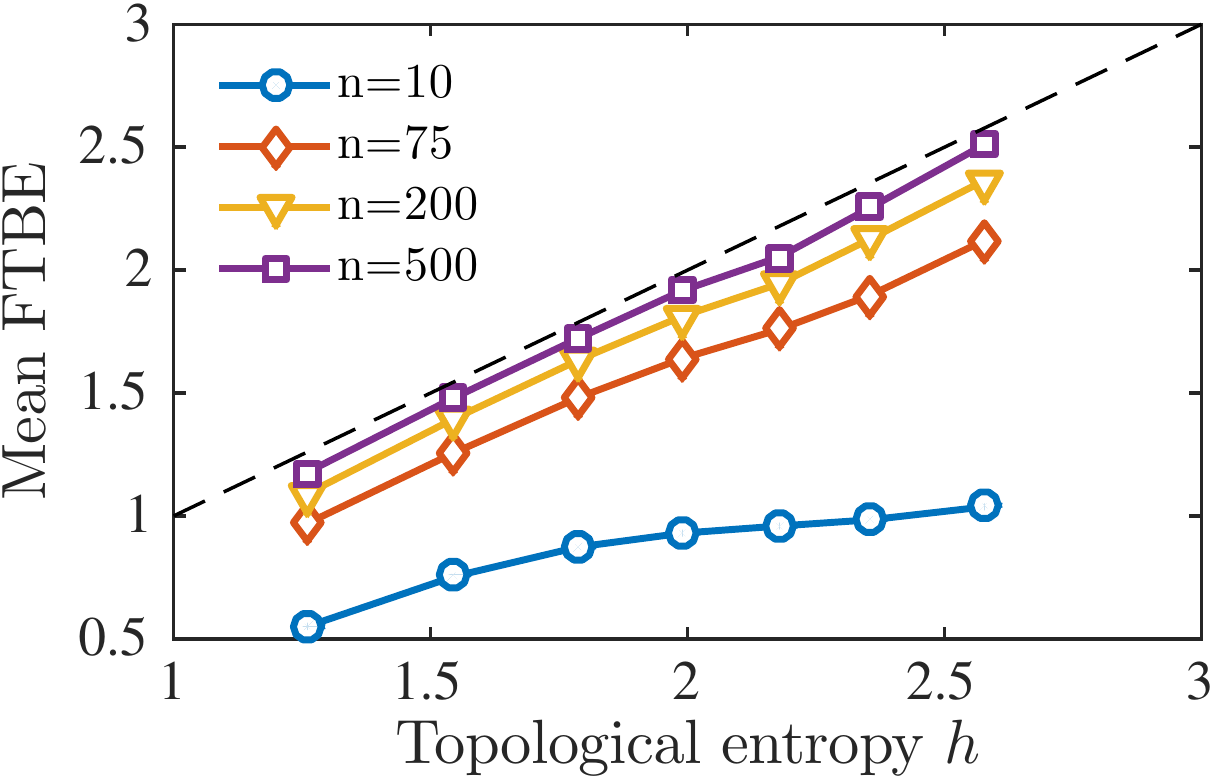}
  \caption{Correlation of topological entropy and mean FTBEs computed using different numbers of strands \(n\). Points were obtained by varying circulation \(\mu\).}\label{fig:ftbe-topentropy-correlation}
\end{figure}

\autoref{fig:ftbe-topentropy-correlation} demonstrates that topological
entropy \(h\) of the flow and mean values of FTBEs are strongly correlated,
therefore the FTBE can serve as a finite-scale proxy for \(h\). To go further,
and extrapolate values of FTBE to obtain the actual value of topological
entropy \(h\), we need that:\begin{inparaenum}[1)]
\item the FTBE indeed approaches topological entropy as \(n\to\infty\), and
\item the growth of \(\ftbe\) can be described by a relatively simple functional model, whose parameters can be estimated.
\end{inparaenum}
Neither of these statements are rigorously known to be true. Previous
numerical results indicate that topological entropy of braids formed by
periodic trajectories approaches the topological entropy of the flow, but
there are no formal verifications~\cite{Finn2007}. The results of
Section~\ref{sec:dependence-on-nstrands} suggest that FTBEs behave similarly.

\begin{figure}[h!]
  \centering
  \includegraphics{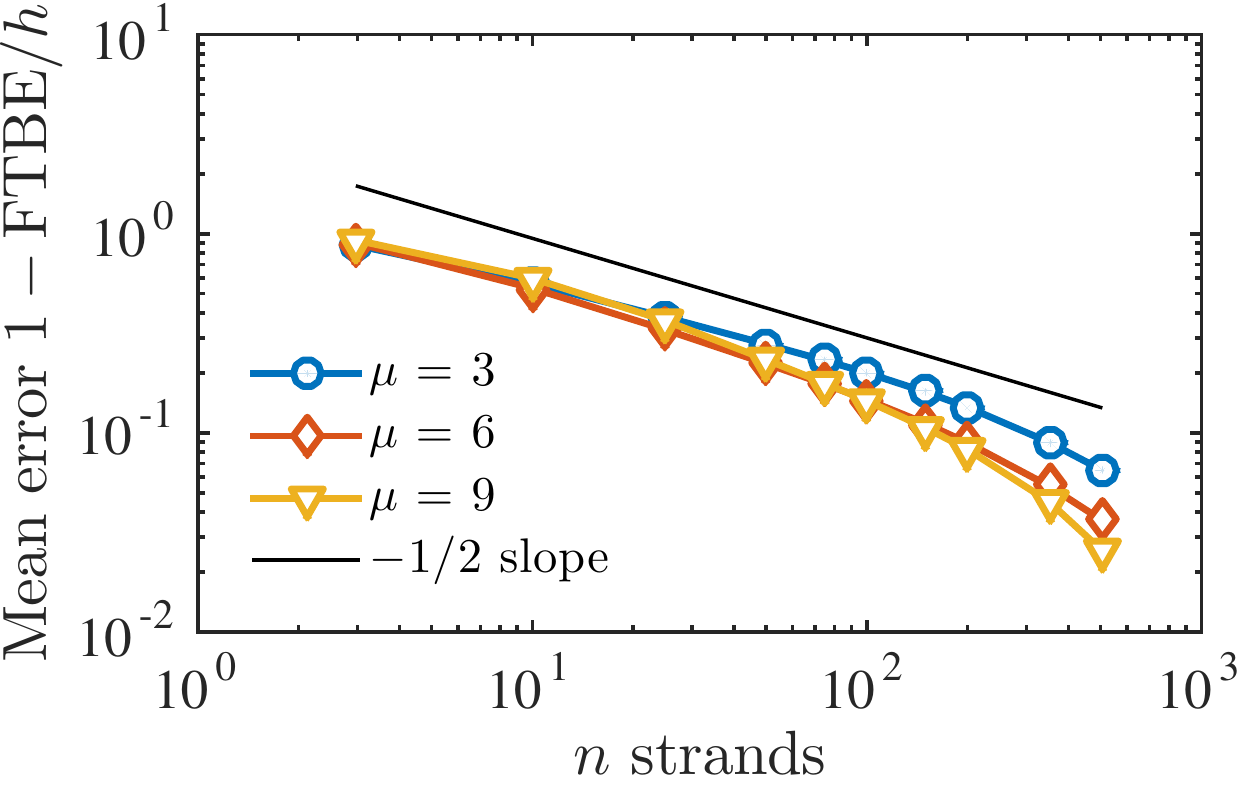}
  \caption{Relative difference of the mean FTBE and topological entropy \(h\)
    for the \arefmodel{}. Statistics are computed with respect to~\(100\)
    \(n\)-\subbraid{}s of a set of 550 trajectories of length
    \hbox{\(T=100\)}.}\label{fig:convergence-to-topent}
\end{figure}

To investigate the approach of FTBEs to topological entropy \(h\), we compute
the relative difference \(1 - \ftbe/h\), and compute its statistics
across sets of \(S = 100\) \subbraid{}s of a physical braid of \(m=550\)
trajectories. \autoref{fig:convergence-to-topent} shows that the relative
difference approximately decays as a power law with slope \(-1/2\) for a large
range of strands \(n\). Nevertheless, the decay deviates from the power law as
\(n\) increases, which means that extrapolating data using a power law model
would overestimate the value of topological entropy.

To estimate the decay of the relative difference more carefully, we fit
two functions to the calculated points:
\begin{description}
\item[Tempered Power Law]
  \begin{equation}
    \label{eq:tpl-error}
    n^{\gamma} e^{\beta n}
  \end{equation}
\item[Log-normal CDF]
  \begin{equation}
    \label{eq:logn-error}
    \erfc \{\gamma \log \beta n\}
  \end{equation}
\end{description}
These two functions are often used as competing statistical models for tails
of distributions, where the criterion for choosing one over the other is based
on a maximum likelihood estimation.~\cite{Clauset2009,Mitzenmacher2004} Since
relative difference at \(n\) strands is not associated with a distribution, we
use the more classical weighted least-squared-error fitting of models to
data. Weights assigned to data points are inversely proportional to the
standard deviation of FTBEs, which enforces a tighter fit to points at larger
\(n\), due to decaying standard deviation, shown in
Figure~\ref{fig:subbraid-comp-ftbe-std}.

\begin{figure}[h!]
  \centering
  \subfigure[\ Fit of functions ~\eqref{eq:tpl-error} and~\eqref{eq:logn-error} (lines) and data (circles) to relative difference between mean FTBE and topological entropy\label{fig:model-curves-error}]{\includegraphics{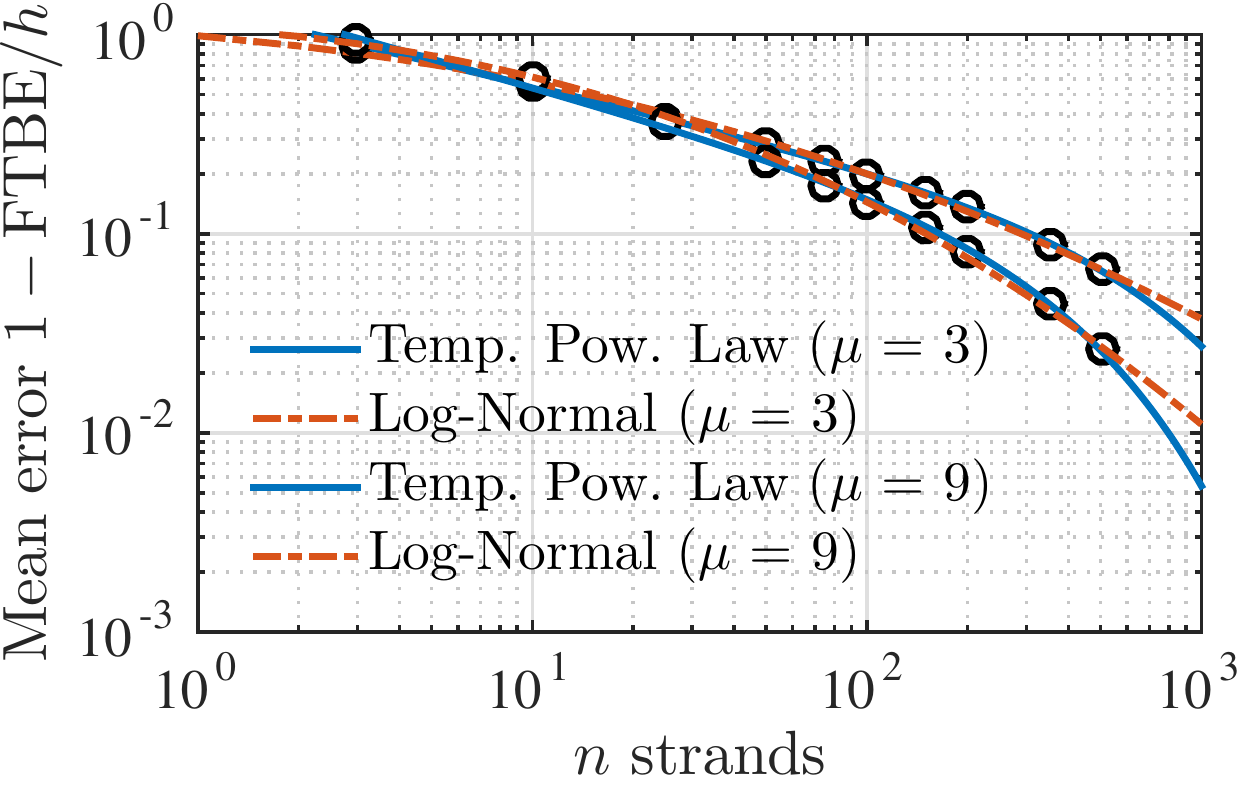}}
  \subfigure[\ Weighted least-squared-error fit residuals\label{fig:wlsq-error}]{\includegraphics{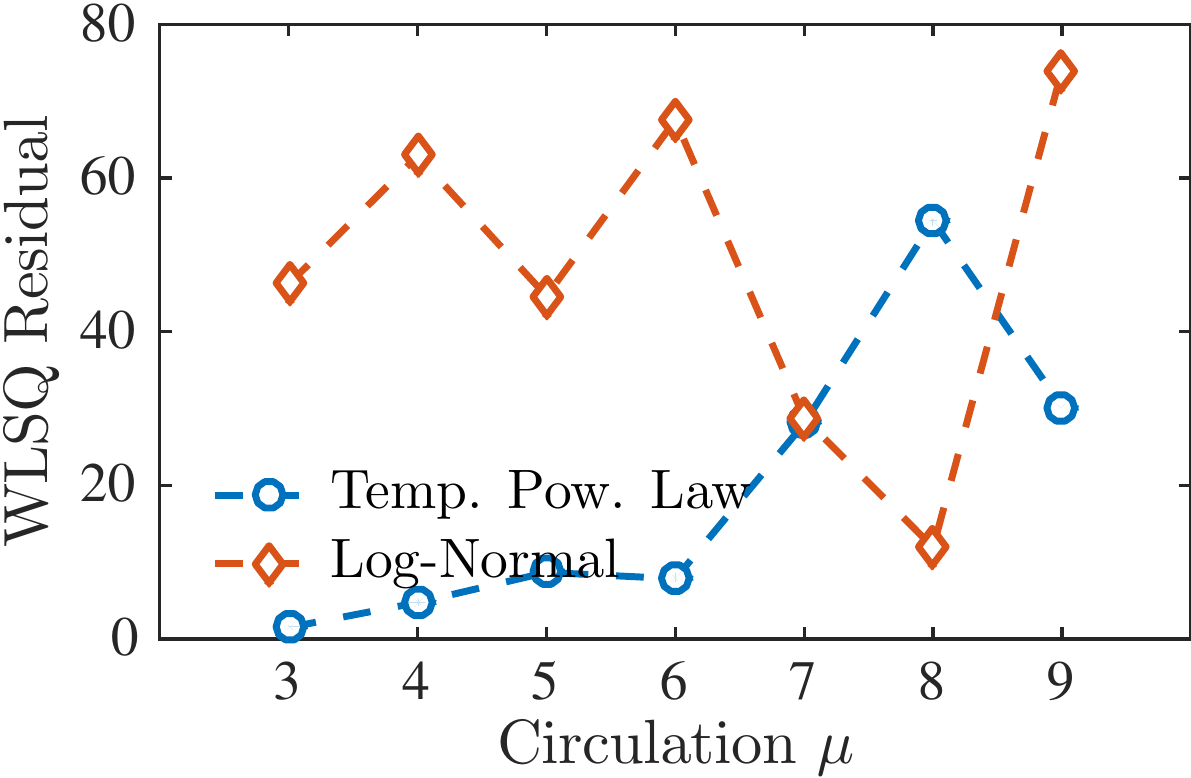}}
  \caption{Quality of fit of Tempered Power Law and Log-normal CDF models to
    computed mean FTBE}\label{fig:quality-of-fit-error}
\end{figure}

Figure~\ref{fig:quality-of-fit-error} shows the quality of fit of the two
models to values of FTBEs. Both functions fit well to data, as evidenced by
Figure~\ref{fig:model-curves-error}. Figure~\ref{fig:wlsq-error} shows
residual errors for both models at different values of circulation and
demonstrates that the tempered power law fits slightly better.

The true test of the models should be their ability to correctly predict the
topological entropy \(h\) as their limit. Instead of using values of \(h\) as
a known limit of proposed models, we introduce a limit parameter
\(h_{\infty}\) and allow it to vary during fitting. This additionally means
that we fit our model functions to FTBE data instead of the relative
difference to a reference value. Other than these modifications, models and
fitting technique remain the same.

\begin{figure}[h!]
  \centering
  \subfigure[\ Tempered Power Law]{\includegraphics{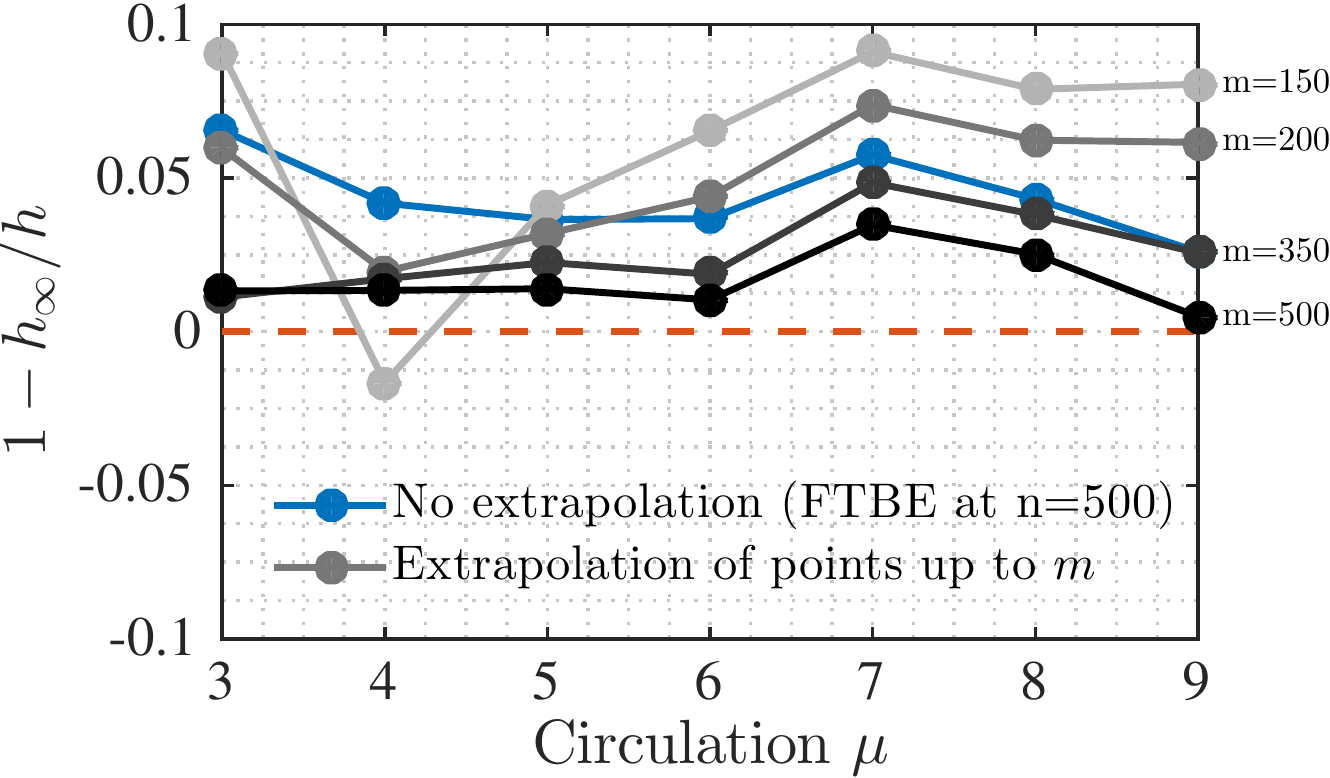}}\hfill
  \subfigure[\ Log Normal CDF]{\includegraphics{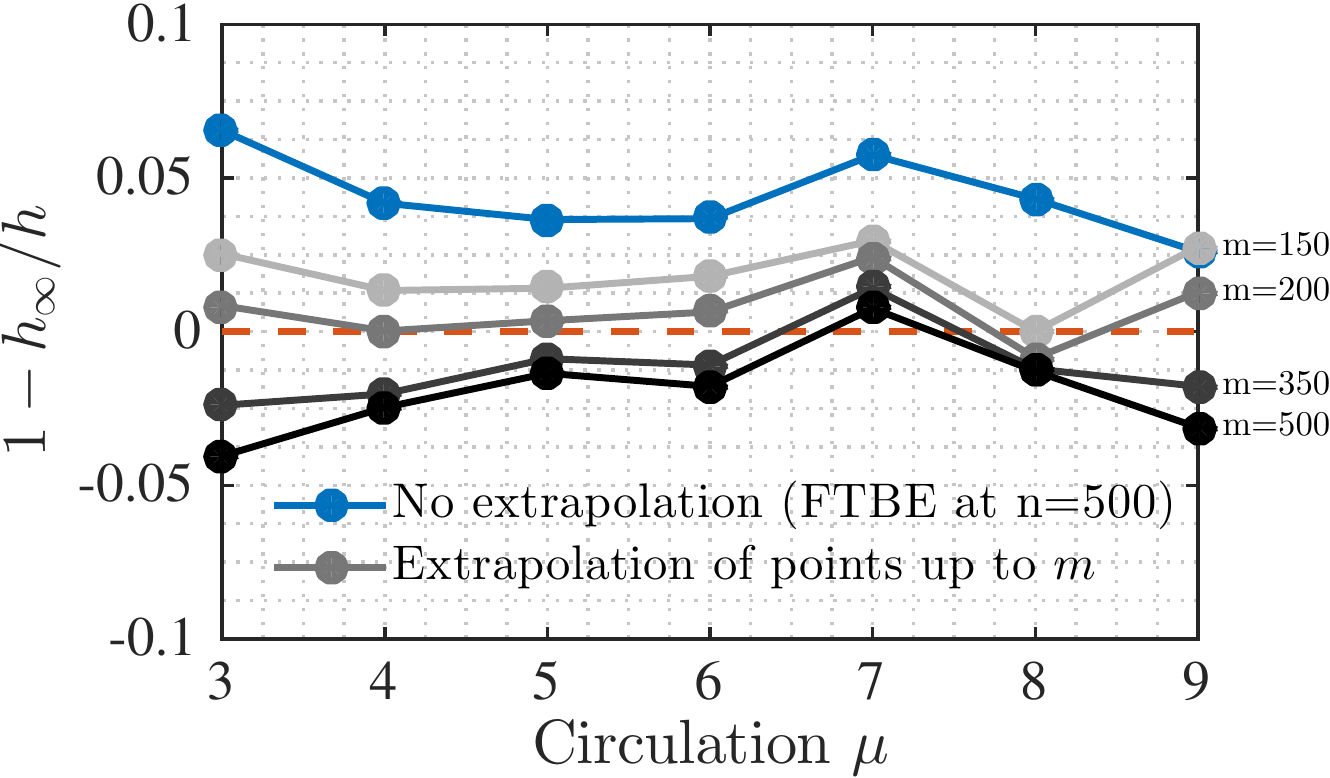}}
  \caption{Comparison of \(h_{\infty}\) obtained by extrapolating tempered
    power law and log-normal models for mean FTBE.  Blue lines show the best estimate of topological entropy obtained without extrapolation. Different gray lines connect values \(h_{\infty}\) obtained from models fit to larger and larger point sets; number next to the line is the number of strands of last point used for fitting.}\label{fig:extrapolation-comparison}
\end{figure}

Unsurprisingly, we again find that both models provide a tight fit to
data; however, their limits differ due to different rates of growth as \(n \to
\infty\). \autoref{fig:extrapolation-comparison} shows relative
differences \(1 - h_{\infty}/h\) between extrapolated limits of FTBE
\(h_{\infty}\) and topological entropy \(h\), depending on circulation
\(\mu\). In addition to fitting models to all available data points, we also
fit them to data sets with points at higher values of \(n\) removed, to test
robustness of the extrapolations \(h_{\infty}\) to the amount of data
used. Only non-negative values of differences are consistent with the
expectation that values of FTBE approach \(h\) from below as \(n\to\infty\).

\autoref{fig:extrapolation-comparison} demonstrates that the Log-normal model
consistently overestimates the value of topological entropy. On the other
hand, extrapolations of the tempered power law provide a better estimate than
the raw FTBE calculation at the largest number of strands (in blue), while
remaining consistent with the theory. We note that extrapolating the fit up to
\(m=350\) strands yields an estimate that is of similar quality to estimation
of \(h\) by FTBE at \(n=500\) strands without extrapolation. This translates
to savings in data collection and analysis time, since converting trajectories
to braids takes \(\mathcal{O}(n^{2})\) time to complete, and computing a single braid at a large \(n\) can take longer than computing many collections of \subbraid{}s at \(m < n\).

The numerical evidence indicates that a tempered power law models the approach
of FTBE to topological entropy, with power law exponents \(\gamma\)
estimated to lie in \([-0.4,-0.5]\). Fixing the exponent of
the tempered power law to \(\gamma=-1/2\) minimally changes quality of fit and
extrapolation by the tempered power law model, suggesting that growth of FTBE
follows an universal behavior, regardless of circulation of the flow.

We caution, however, against making stronger arguments based on the
experiments performed, because the reference value for topological entropy
\(h\) is an estimate itself, computed by material advection. Since the
precision achieved by FTBE is within 1\% to 5\% of \(h\), it is likely that it
approaches the confidence interval for the estimation by material
advection. For quantitative conclusions about the approach of FTBE to
topological entropy of the flow, analytical justification for a model of
growth of FTBE is needed.

\section{Discussion}
\label{sec:discussion}

We briefly summarize our findings about Finite-Time Braiding Exponents.
\begin{compactitem}
\item FTBEs are very robust with respect to the projection angle \(\alpha\)
  used to construct the braids.
\item Reasonably-small values of the time step \(\tau\) lead to well-defined
  braids, but FTBEs may tolerate a larger \(\tau\).
\item Increasing the length of trajectories in mixing dynamics results in
  convergence of FTBEs. At the same time, the spatial mean of FTBEs
  stabilizes, while their spatial fluctuations decay.
\item The spatial mean of FTBEs correlates strongly with the topological
  entropy of the flow; as the number of strands \(n\) is increased, the mean
  approaches the topological entropy.
\end{compactitem}

The Finite-Time Braiding Exponent of \(n\) trajectories sampled from the flow
can be interpreted as a ``multi-scale'' measure of complexity of the flow at
scales determined by \(n\). An important finding is that to estimate
complexity at different scales, a single collection of \(n\) trajectories can
be re-used by sampling \subbraid{}s. Even though such \subbraid{}s are not
fully independent, since a given trajectory will participate in many braids,
the procedure yields statistics of FTBEs similar in quality to those computed
using independently-sampled braids. This technique greatly reduces the
required amount of input data without sacrificing the quality of computed FTBEs
at various scales.

The value of FTBEs seems to approach the topological entropy of the flow as
\(n\) is increased, but more work is needed to establish the asymptotics of
that approach. Fitting plausible functional models suggests that for
trajectories of mixing dynamical systems the difference between FTBEs and
topological entropy decays according to a power law tempered by an
exponential. If confirmed analytically, this relation would allow one to
estimate free parameters of the model from values of FTBEs at several
different scales \(n\), and extrapolate the model as \(n\to\infty\) to
approximate the topological entropy very closely.

While these conclusions are based on the study of the \arefmodel{}, it is
reasonable to expect similar results for other mixing flows. In flows that are
not spatially homogeneous, e.g., those containing coherent substructures,
lengths of strands and locations of their initial conditions will be of
crucial importance for interpretation of FTBE values. In such situations,
other quantities based on braids can be computed to provide a more complete
picture of dynamics. In particular, a promising approach may be to couple
braid approximations of coherent structures\cite{Allshouse2012} with FTBE
estimates of complexity inside those structures to assess the spatial
distribution of different dynamics in more complicated flows, even when the
only information available is a finite set of trajectories.

\acknowledgments{}

The authors thank Michael Allshouse, Margaux Filippi, and Tom Peacock for
helpful discussions.  This research was supported by the US National Science
Foundation, under grant CMMI-1233935.

\bibliography{ftbe-paper}

%merlin.mbs aipnum4-1.bst 2010-07-25 4.21a (PWD, AO, DPC) hacked
%Control: key (0)
%Control: author (8) initials jnrlst
%Control: editor formatted (1) identically to author
%Control: production of article title (0) allowed
%Control: page (1) range
%Control: year (1) truncated
%Control: production of eprint (0) enabled
\begin{thebibliography}{66}%
\makeatletter
\providecommand \@ifxundefined [1]{%
 \@ifx{#1\undefined}
}%
\providecommand \@ifnum [1]{%
 \ifnum #1\expandafter \@firstoftwo
 \else \expandafter \@secondoftwo
 \fi
}%
\providecommand \@ifx [1]{%
 \ifx #1\expandafter \@firstoftwo
 \else \expandafter \@secondoftwo
 \fi
}%
\providecommand \natexlab [1]{#1}%
\providecommand \enquote  [1]{``#1''}%
\providecommand \bibnamefont  [1]{#1}%
\providecommand \bibfnamefont [1]{#1}%
\providecommand \citenamefont [1]{#1}%
\providecommand \href@noop [0]{\@secondoftwo}%
\providecommand \href [0]{\begingroup \@sanitize@url \@href}%
\providecommand \@href[1]{\@@startlink{#1}\@@href}%
\providecommand \@@href[1]{\endgroup#1\@@endlink}%
\providecommand \@sanitize@url [0]{\catcode `\\12\catcode `\$12\catcode
  `\&12\catcode `\#12\catcode `\^12\catcode `\_12\catcode `\%12\relax}%
\providecommand \@@startlink[1]{}%
\providecommand \@@endlink[0]{}%
\providecommand \url  [0]{\begingroup\@sanitize@url \@url }%
\providecommand \@url [1]{\endgroup\@href {#1}{\urlprefix }}%
\providecommand \urlprefix  [0]{URL }%
\providecommand \Eprint [0]{\href }%
\providecommand \doibase [0]{http://dx.doi.org/}%
\providecommand \selectlanguage [0]{\@gobble}%
\providecommand \bibinfo  [0]{\@secondoftwo}%
\providecommand \bibfield  [0]{\@secondoftwo}%
\providecommand \translation [1]{[#1]}%
\providecommand \BibitemOpen [0]{}%
\providecommand \bibitemStop [0]{}%
\providecommand \bibitemNoStop [0]{.\EOS\space}%
\providecommand \EOS [0]{\spacefactor3000\relax}%
\providecommand \BibitemShut  [1]{\csname bibitem#1\endcsname}%
\let\auto@bib@innerbib\@empty
%</preamble>
\bibitem [{\citenamefont {Aref}(1984)}]{Aref1984}%
  \BibitemOpen
  \bibfield  {author} {\bibinfo {author} {\bibfnamefont {H.}~\bibnamefont
  {Aref}},\ }\bibfield  {title} {\enquote {\bibinfo {title} {Stirring by
  chaotic advection},}\ }\href {\doibase 10.1017/S0022112084001233} {\bibfield
  {journal} {\bibinfo  {journal} {Journal of {Fluid} {Mechanics}}\ }\textbf
  {\bibinfo {volume} {143}},\ \bibinfo {pages} {1--21} (\bibinfo {year}
  {1984})}\BibitemShut {NoStop}%
\bibitem [{\citenamefont {Ottino}(1989)}]{Ottino1989}%
  \BibitemOpen
  \bibfield  {author} {\bibinfo {author} {\bibfnamefont {J.~M.}\ \bibnamefont
  {Ottino}},\ }\href {http://www.cambridge.org/9780521368780} {\emph {\bibinfo
  {title} {The kinematics of mixing: stretching, chaos, and transport}}},\
  Cambridge {Texts} in {Applied} {Mathematics}\ (\bibinfo  {publisher}
  {Cambridge {University} {Press}},\ \bibinfo {address} {Cambridge},\ \bibinfo
  {year} {1989})\BibitemShut {NoStop}%
\bibitem [{\citenamefont {Young}(2003)}]{Young2003a}%
  \BibitemOpen
  \bibfield  {author} {\bibinfo {author} {\bibfnamefont {L.-S.}\ \bibnamefont
  {Young}},\ }\bibfield  {title} {\enquote {\bibinfo {title} {Entropy in
  dynamical systems},}\ }in\ \href
  {http://www.ams.org/mathscinet-getitem?mr=2035829} {\emph {\bibinfo
  {booktitle} {Entropy}}},\ \bibinfo {series and number} {Princeton {Ser}.
  {Appl}. {Math}.}\ (\bibinfo  {publisher} {Princeton {Univ}. {Press},
  {Princeton}, {NJ}},\ \bibinfo {year} {2003})\ pp.\ \bibinfo {pages}
  {313--327}\BibitemShut {NoStop}%
\bibitem [{\citenamefont {Bollt}\ and\ \citenamefont
  {Santitissadeekorn}(2013)}]{Bollt2013}%
  \BibitemOpen
  \bibfield  {author} {\bibinfo {author} {\bibfnamefont {E.~M.}\ \bibnamefont
  {Bollt}}\ and\ \bibinfo {author} {\bibfnamefont {N.}~\bibnamefont
  {Santitissadeekorn}},\ }\href
  {http://www.ams.org/mathscinet-getitem?mr=3154435} {\emph {\bibinfo {title}
  {Applied and computational measurable dynamics}}},\ \bibinfo {series}
  {Mathematical {Modeling} and {Computation}}, Vol.~\bibinfo {volume} {18}\
  (\bibinfo  {publisher} {Society for {Industrial} and {Applied} {Mathematics}
  ({SIAM}), {Philadelphia}, {PA}},\ \bibinfo {year} {2013})\BibitemShut
  {NoStop}%
\bibitem [{\citenamefont {Haszpra}\ and\ \citenamefont
  {Tel}(2013)}]{Haszpra2013a}%
  \BibitemOpen
  \bibfield  {author} {\bibinfo {author} {\bibfnamefont {T.}~\bibnamefont
  {Haszpra}}\ and\ \bibinfo {author} {\bibfnamefont {T.}~\bibnamefont {Tel}},\
  }\bibfield  {title} {\enquote {\bibinfo {title} {Topological {Entropy}: {A}
  {Lagrangian} {Measure} of the {State} of the {Free} {Atmosphere}},}\ }\href
  {\doibase 10.1175/JAS-D-13-069.1} {\bibfield  {journal} {\bibinfo  {journal}
  {Journal of the {Atmospheric} {Sciences}}\ }\textbf {\bibinfo {volume}
  {70}},\ \bibinfo {pages} {4030--4040} (\bibinfo {year} {2013})}\BibitemShut
  {NoStop}%
\bibitem [{\citenamefont {Bowen}(1978)}]{Bowen1978}%
  \BibitemOpen
  \bibfield  {author} {\bibinfo {author} {\bibfnamefont {R.}~\bibnamefont
  {Bowen}},\ }\bibfield  {title} {\enquote {\bibinfo {title} {Entropy and the
  fundamental group},}\ }in\ \href
  {http://www.ams.org/mathscinet-getitem?mr=518545} {\emph {\bibinfo
  {booktitle} {The structure of attractors in dynamical systems ({Proc}.
  {Conf}., {North} {Dakota} {State} {Univ}., {Fargo}, {N}.{D}., 1977)}}},\
  \bibinfo {series} {Lecture {Notes} in {Math}.}, Vol.\ \bibinfo {volume}
  {668}\ (\bibinfo  {publisher} {Springer, {Berlin}},\ \bibinfo {year} {1978})\
  pp.\ \bibinfo {pages} {21--29}\BibitemShut {NoStop}%
\bibitem [{\citenamefont {Froyland}, \citenamefont {Junge},\ and\ \citenamefont
  {Ochs}(2001)}]{Froyland2001}%
  \BibitemOpen
  \bibfield  {author} {\bibinfo {author} {\bibfnamefont {G.}~\bibnamefont
  {Froyland}}, \bibinfo {author} {\bibfnamefont {O.}~\bibnamefont {Junge}}, \
  and\ \bibinfo {author} {\bibfnamefont {G.}~\bibnamefont {Ochs}},\ }\bibfield
  {title} {\enquote {\bibinfo {title} {Rigorous computation of topological
  entropy with respect to a finite partition},}\ }\href {\doibase
  10.1016/S0167-2789(01)00216-0} {\bibfield  {journal} {\bibinfo  {journal}
  {Physica {D}. {Nonlinear} {Phenomena}}\ }\textbf {\bibinfo {volume} {154}},\
  \bibinfo {pages} {68--84} (\bibinfo {year} {2001})}\BibitemShut {NoStop}%
\bibitem [{\citenamefont {Froyland}\ and\ \citenamefont
  {Padberg-Gehle}(2012)}]{Froyland2012a}%
  \BibitemOpen
  \bibfield  {author} {\bibinfo {author} {\bibfnamefont {G.}~\bibnamefont
  {Froyland}}\ and\ \bibinfo {author} {\bibfnamefont {K.}~\bibnamefont
  {Padberg-Gehle}},\ }\bibfield  {title} {\enquote {\bibinfo {title}
  {Finite-time entropy: {A} probabilistic approach for measuring nonlinear
  stretching},}\ }\href {\doibase 10.1016/j.physd.2012.06.010} {\bibfield
  {journal} {\bibinfo  {journal} {Physica {D}: {Nonlinear} {Phenomena}}\
  }\textbf {\bibinfo {volume} {241}},\ \bibinfo {pages} {1612--1628} (\bibinfo
  {year} {2012})}\BibitemShut {NoStop}%
\bibitem [{\citenamefont {D'Alessandro}, \citenamefont {Dahleh},\ and\
  \citenamefont {Mezic}(1999)}]{DAlessandro1999a}%
  \BibitemOpen
  \bibfield  {author} {\bibinfo {author} {\bibfnamefont {D.}~\bibnamefont
  {D'Alessandro}}, \bibinfo {author} {\bibfnamefont {M.}~\bibnamefont
  {Dahleh}}, \ and\ \bibinfo {author} {\bibfnamefont {I.}~\bibnamefont
  {Mezic}},\ }\bibfield  {title} {\enquote {\bibinfo {title} {Control of mixing
  in fluid flow: a maximum entropy approach},}\ }\href {\doibase
  10.1109/9.793724} {\bibfield  {journal} {\bibinfo  {journal} {{IEEE}
  {Transactions} on {Automatic} {Control}}\ }\textbf {\bibinfo {volume} {44}},\
  \bibinfo {pages} {1852--1863} (\bibinfo {year} {1999})}\BibitemShut {NoStop}%
\bibitem [{\citenamefont {Katok}(1980)}]{Katok1980}%
  \BibitemOpen
  \bibfield  {author} {\bibinfo {author} {\bibfnamefont {A.}~\bibnamefont
  {Katok}},\ }\bibfield  {title} {\enquote {\bibinfo {title} {Lyapunov
  exponents, entropy and periodic orbits for diffeomorphisms},}\ }\href
  {http://www.ams.org/mathscinet-getitem?mr=573822} {\bibfield  {journal}
  {\bibinfo  {journal} {Institut des {Hautes} \'{E}tudes {Scientifiques}.
  {Publications} {Math}\'{e}matiques}\ }\textbf {\bibinfo {volume} {51}},\
  \bibinfo {pages} {137--173} (\bibinfo {year} {1980})}\BibitemShut {NoStop}%
\bibitem [{\citenamefont {Waugh}, \citenamefont {Keating},\ and\ \citenamefont
  {Chen}(2012)}]{Waugh2012}%
  \BibitemOpen
  \bibfield  {author} {\bibinfo {author} {\bibfnamefont {D.~W.}\ \bibnamefont
  {Waugh}}, \bibinfo {author} {\bibfnamefont {S.~R.}\ \bibnamefont {Keating}},
  \ and\ \bibinfo {author} {\bibfnamefont {M.-L.}\ \bibnamefont {Chen}},\
  }\bibfield  {title} {\enquote {\bibinfo {title} {Diagnosing {Ocean}
  {Stirring}: {Comparison} of {Relative} {Dispersion} and {Finite}-{Time}
  {Lyapunov} {Exponents}},}\ }\href {\doibase 10.1175/JPO-D-11-0215.1}
  {\bibfield  {journal} {\bibinfo  {journal} {Journal of {Physical}
  {Oceanography}}\ }\textbf {\bibinfo {volume} {42}},\ \bibinfo {pages}
  {1173--1185} (\bibinfo {year} {2012})}\BibitemShut {NoStop}%
\bibitem [{\citenamefont {Marshall}\ \emph {et~al.}(2006)\citenamefont
  {Marshall}, \citenamefont {Shuckburgh}, \citenamefont {Jones},\ and\
  \citenamefont {Hill}}]{Marshall2006}%
  \BibitemOpen
  \bibfield  {author} {\bibinfo {author} {\bibfnamefont {J.}~\bibnamefont
  {Marshall}}, \bibinfo {author} {\bibfnamefont {E.}~\bibnamefont
  {Shuckburgh}}, \bibinfo {author} {\bibfnamefont {H.}~\bibnamefont {Jones}}, \
  and\ \bibinfo {author} {\bibfnamefont {C.}~\bibnamefont {Hill}},\ }\bibfield
  {title} {\enquote {\bibinfo {title} {Estimates and implications of surface
  eddy diffusivity in the {Southern} {Ocean} derived from tracer transport},}\
  }\href {\doibase 10.1175/JPO2949.1} {\bibfield  {journal} {\bibinfo
  {journal} {Journal of {Physical} {Oceanography}}\ }\textbf {\bibinfo {volume}
  {36}},\ \bibinfo {pages} {1806--1821} (\bibinfo {year} {2006})}\BibitemShut
  {NoStop}%
\bibitem [{\citenamefont {Pierrehumbert}(1991)}]{Pierrehumbert1991}%
  \BibitemOpen
  \bibfield  {author} {\bibinfo {author} {\bibfnamefont {R.}~\bibnamefont
  {Pierrehumbert}},\ }\bibfield  {title} {\enquote {\bibinfo {title}
  {Large-{Scale} {Horizontal} {Mixing} in {Planetary}-{Atmospheres}},}\ }\href
  {\doibase 10.1063/1.858053} {\bibfield  {journal} {\bibinfo  {journal}
  {Physics of {Fluids} {A}--{Fluid} {Dynamics}}\ }\textbf {\bibinfo {volume}
  {3}},\ \bibinfo {pages} {1250--1260} (\bibinfo {year} {1991})}\BibitemShut
  {NoStop}%
\bibitem [{\citenamefont {Aurell}\ \emph {et~al.}(1997)\citenamefont {Aurell},
  \citenamefont {Boffetta}, \citenamefont {Crisanti}, \citenamefont {Paladin},\
  and\ \citenamefont {Vulpiani}}]{Aurell1997}%
  \BibitemOpen
  \bibfield  {author} {\bibinfo {author} {\bibfnamefont {E.}~\bibnamefont
  {Aurell}}, \bibinfo {author} {\bibfnamefont {G.}~\bibnamefont {Boffetta}},
  \bibinfo {author} {\bibfnamefont {A.}~\bibnamefont {Crisanti}}, \bibinfo
  {author} {\bibfnamefont {G.}~\bibnamefont {Paladin}}, \ and\ \bibinfo
  {author} {\bibfnamefont {A.}~\bibnamefont {Vulpiani}},\ }\bibfield  {title}
  {\enquote {\bibinfo {title} {Predictability in the large: {An} extension of
  the concept of {Lyapunov} exponent},}\ }\href {\doibase
  10.1088/0305-4470/30/1/003} {\bibfield  {journal} {\bibinfo  {journal}
  {Journal of {Physics} a-{Mathematical} and {General}}\ }\textbf {\bibinfo
  {volume} {30}},\ \bibinfo {pages} {1--26} (\bibinfo {year}
  {1997})}\BibitemShut {NoStop}%
\bibitem [{\citenamefont {Artale}\ \emph {et~al.}(1997)\citenamefont {Artale},
  \citenamefont {Boffetta}, \citenamefont {Celani}, \citenamefont {Cencini},\
  and\ \citenamefont {Vulpiani}}]{Artale1997}%
  \BibitemOpen
  \bibfield  {author} {\bibinfo {author} {\bibfnamefont {V.}~\bibnamefont
  {Artale}}, \bibinfo {author} {\bibfnamefont {G.}~\bibnamefont {Boffetta}},
  \bibinfo {author} {\bibfnamefont {A.}~\bibnamefont {Celani}}, \bibinfo
  {author} {\bibfnamefont {M.}~\bibnamefont {Cencini}}, \ and\ \bibinfo
  {author} {\bibfnamefont {A.}~\bibnamefont {Vulpiani}},\ }\bibfield  {title}
  {\enquote {\bibinfo {title} {Dispersion of passive tracers in closed basins:
  {Beyond} the diffusion coefficient},}\ }\href {\doibase 10.1063/1.869433}
  {\bibfield  {journal} {\bibinfo  {journal} {Physics of {Fluids}}\ }\textbf
  {\bibinfo {volume} {9}},\ \bibinfo {pages} {3162--3171} (\bibinfo {year}
  {1997})}\BibitemShut {NoStop}%
\bibitem [{\citenamefont {Joseph}\ and\ \citenamefont
  {Legras}(2002)}]{Joseph2002}%
  \BibitemOpen
  \bibfield  {author} {\bibinfo {author} {\bibfnamefont {B.}~\bibnamefont
  {Joseph}}\ and\ \bibinfo {author} {\bibfnamefont {B.}~\bibnamefont
  {Legras}},\ }\bibfield  {title} {\enquote {\bibinfo {title} {Relation between
  kinematic boundaries, stirring, and barriers for the {Antarctic} polar
  vortex},}\ }\href {\doibase 10.1175/1520-0469(2002)059<1198:RBKBSA>2.0.CO;2}
  {\bibfield  {journal} {\bibinfo  {journal} {Journal of the {Atmospheric}
  {Sciences}}\ }\textbf {\bibinfo {volume} {59}},\ \bibinfo {pages}
  {1198--1212} (\bibinfo {year} {2002})}\BibitemShut {NoStop}%
\bibitem [{\citenamefont {d'Ovidio}\ \emph {et~al.}(2004)\citenamefont
  {d'Ovidio}, \citenamefont {Fernandez}, \citenamefont {Hernandez-Garcia},\
  and\ \citenamefont {Lopez}}]{dOvidio2004}%
  \BibitemOpen
  \bibfield  {author} {\bibinfo {author} {\bibfnamefont {F.}~\bibnamefont
  {d'Ovidio}}, \bibinfo {author} {\bibfnamefont {V.}~\bibnamefont {Fernandez}},
  \bibinfo {author} {\bibfnamefont {E.}~\bibnamefont {Hernandez-Garcia}}, \
  and\ \bibinfo {author} {\bibfnamefont {C.}~\bibnamefont {Lopez}},\ }\bibfield
   {title} {\enquote {\bibinfo {title} {Mixing structures in the
  {Mediterranean} {Sea} from finite-size {Lyapunov} exponents},}\ }\href
  {\doibase 10.1029/2004GL020328} {\bibfield  {journal} {\bibinfo  {journal}
  {Geophysical {Research} {Letters}}\ }\textbf {\bibinfo {volume} {31}},\
  \bibinfo {pages} {L17203} (\bibinfo {year} {2004})}\BibitemShut {NoStop}%
\bibitem [{\citenamefont {Farnetani}\ and\ \citenamefont
  {Samuel}(2003)}]{Farnetani2003}%
  \BibitemOpen
  \bibfield  {author} {\bibinfo {author} {\bibfnamefont {C.~G.}\ \bibnamefont
  {Farnetani}}\ and\ \bibinfo {author} {\bibfnamefont {H.}~\bibnamefont
  {Samuel}},\ }\bibfield  {title} {\enquote {\bibinfo {title} {Lagrangian
  structures and stirring in the {Earth}'s mantle},}\ }\href {\doibase
  10.1016/S0012-821X(02)01085-3} {\bibfield  {journal} {\bibinfo  {journal}
  {Earth and {Planetary} {Science} {Letters}}\ }\textbf {\bibinfo {volume}
  {206}},\ \bibinfo {pages} {335--348} (\bibinfo {year} {2003})}\BibitemShut
  {NoStop}%
\bibitem [{\citenamefont {Haller}(2002)}]{Haller2002}%
  \BibitemOpen
  \bibfield  {author} {\bibinfo {author} {\bibfnamefont {G.}~\bibnamefont
  {Haller}},\ }\bibfield  {title} {\enquote {\bibinfo {title} {Lagrangian
  coherent structures from approximate velocity data},}\ }\href {\doibase
  10.1063/1.1477449} {\bibfield  {journal} {\bibinfo  {journal} {Physics of
  {Fluids}}\ }\textbf {\bibinfo {volume} {14}},\ \bibinfo {pages} {1851--1861}
  (\bibinfo {year} {2002})}\BibitemShut {NoStop}%
\bibitem [{\citenamefont {Haller}(2001)}]{Haller2001a}%
  \BibitemOpen
  \bibfield  {author} {\bibinfo {author} {\bibfnamefont {G.}~\bibnamefont
  {Haller}},\ }\bibfield  {title} {\enquote {\bibinfo {title} {Distinguished
  material surfaces and coherent structures in three-dimensional fluid
  flows},}\ }\href {\doibase 10.1016/S0167-2789(00)00199-8} {\bibfield
  {journal} {\bibinfo  {journal} {Physica {D}: {Nonlinear} {Phenomena}}\
  }\textbf {\bibinfo {volume} {149}},\ \bibinfo {pages} {248--277} (\bibinfo
  {year} {2001})}\BibitemShut {NoStop}%
\bibitem [{\citenamefont {Haller}\ and\ \citenamefont
  {Yuan}(2000)}]{Haller2000}%
  \BibitemOpen
  \bibfield  {author} {\bibinfo {author} {\bibfnamefont {G.}~\bibnamefont
  {Haller}}\ and\ \bibinfo {author} {\bibfnamefont {G.}~\bibnamefont {Yuan}},\
  }\bibfield  {title} {\enquote {\bibinfo {title} {Lagrangian coherent
  structures and mixing in two-dimensional turbulence},}\ }\href {\doibase
  10.1016/S0167-2789(00)00142-1} {\bibfield  {journal} {\bibinfo  {journal}
  {Physica {D}. {Nonlinear} {Phenomena}}\ }\textbf {\bibinfo {volume} {147}},\
  \bibinfo {pages} {352--370} (\bibinfo {year} {2000})}\BibitemShut {NoStop}%
\bibitem [{\citenamefont {Shadden}, \citenamefont {Lekien},\ and\ \citenamefont
  {Marsden}(2005)}]{Shadden2005}%
  \BibitemOpen
  \bibfield  {author} {\bibinfo {author} {\bibfnamefont {S.~C.}\ \bibnamefont
  {Shadden}}, \bibinfo {author} {\bibfnamefont {F.}~\bibnamefont {Lekien}}, \
  and\ \bibinfo {author} {\bibfnamefont {J.~E.}\ \bibnamefont {Marsden}},\
  }\bibfield  {title} {\enquote {\bibinfo {title} {Definition and properties of
  {Lagrangian} coherent structures from finite-time {Lyapunov} exponents in
  two-dimensional aperiodic flows},}\ }\href {\doibase
  10.1016/j.physd.2005.10.007} {\bibfield  {journal} {\bibinfo  {journal}
  {Physica {D}. {Nonlinear} {Phenomena}}\ }\textbf {\bibinfo {volume} {212}},\
  \bibinfo {pages} {271--304} (\bibinfo {year} {2005})}\BibitemShut {NoStop}%
\bibitem [{\citenamefont {Haller}\ and\ \citenamefont
  {Sapsis}(2011)}]{Haller2011}%
  \BibitemOpen
  \bibfield  {author} {\bibinfo {author} {\bibfnamefont {G.}~\bibnamefont
  {Haller}}\ and\ \bibinfo {author} {\bibfnamefont {T.}~\bibnamefont
  {Sapsis}},\ }\bibfield  {title} {\enquote {\bibinfo {title} {Lagrangian
  coherent structures and the smallest finite-time {Lyapunov} exponent},}\
  }\href {\doibase 10.1063/1.3579597} {\bibfield  {journal} {\bibinfo
  {journal} {Chaos: {An} {Interdisciplinary} {Journal} of {Nonlinear}
  {Science}}\ }\textbf {\bibinfo {volume} {21}},\ \bibinfo {pages} {023115}
  (\bibinfo {year} {2011})}\BibitemShut {NoStop}%
\bibitem [{\citenamefont {Haller}(2015)}]{Haller2015}%
  \BibitemOpen
  \bibfield  {author} {\bibinfo {author} {\bibfnamefont {G.}~\bibnamefont
  {Haller}},\ }\bibfield  {title} {\enquote {\bibinfo {title} {Langrangian
  {Coherent} {Structures}},}\ }\href {\doibase
  10.1146/annurev-fluid-010313-141322} {\bibfield  {journal} {\bibinfo
  {journal} {Annual {Review} of {Fluid} {Mechanics}}\ }\textbf {\bibinfo
  {volume} {47}},\ \bibinfo {pages} {null} (\bibinfo {year}
  {2015})}\BibitemShut {NoStop}%
\bibitem [{\citenamefont {BozorgMagham}, \citenamefont {Ross},\ and\
  \citenamefont {Schmale~III}(2013)}]{BozorgMagham2013}%
  \BibitemOpen
  \bibfield  {author} {\bibinfo {author} {\bibfnamefont {A.~E.}\ \bibnamefont
  {BozorgMagham}}, \bibinfo {author} {\bibfnamefont {S.~D.}\ \bibnamefont
  {Ross}}, \ and\ \bibinfo {author} {\bibfnamefont {D.~G.}\ \bibnamefont
  {Schmale~III}},\ }\bibfield  {title} {\enquote {\bibinfo {title} {Real-time
  prediction of atmospheric {Lagrangian} coherent structures based on forecast
  data: {An} application and error analysis},}\ }\href {\doibase
  10.1016/j.physd.2013.05.003} {\bibfield  {journal} {\bibinfo  {journal}
  {Physica {D}: {Nonlinear} {Phenomena}}\ }\textbf {\bibinfo {volume} {258}},\
  \bibinfo {pages} {47--60} (\bibinfo {year} {2013})}\BibitemShut {NoStop}%
\bibitem [{\citenamefont {Karrasch}\ and\ \citenamefont
  {Haller}(2013)}]{Karrasch2013}%
  \BibitemOpen
  \bibfield  {author} {\bibinfo {author} {\bibfnamefont {D.}~\bibnamefont
  {Karrasch}}\ and\ \bibinfo {author} {\bibfnamefont {G.}~\bibnamefont
  {Haller}},\ }\bibfield  {title} {\enquote {\bibinfo {title} {Do
  {Finite}-{Size} {Lyapunov} {Exponents} detect coherent structures?}}\ }\href
  {\doibase 10.1063/1.4837075} {\bibfield  {journal} {\bibinfo  {journal}
  {Chaos: {An} {Interdisciplinary} {Journal} of {Nonlinear} {Science}}\
  }\textbf {\bibinfo {volume} {23}},\ \bibinfo {pages} {043126} (\bibinfo
  {year} {2013})}\BibitemShut {NoStop}%
\bibitem [{\citenamefont {Samelson}(2013)}]{Samelson2013}%
  \BibitemOpen
  \bibfield  {author} {\bibinfo {author} {\bibfnamefont {R.}~\bibnamefont
  {Samelson}},\ }\bibfield  {title} {\enquote {\bibinfo {title} {Lagrangian
  {Motion}, {Coherent} {Structures}, and {Lines} of {Persistent} {Material}
  {Strain}},}\ }\href {\doibase 10.1146/annurev-marine-120710-100819}
  {\bibfield  {journal} {\bibinfo  {journal} {Annual {Review} of {Marine}
  {Science}}\ }\textbf {\bibinfo {volume} {5}},\ \bibinfo {pages} {137--163}
  (\bibinfo {year} {2013})}\BibitemShut {NoStop}%
\bibitem [{\citenamefont {LaCasce}(2008)}]{LaCasce2008}%
  \BibitemOpen
  \bibfield  {author} {\bibinfo {author} {\bibfnamefont {J.~H.}\ \bibnamefont
  {LaCasce}},\ }\bibfield  {title} {\enquote {\bibinfo {title} {Statistics from
  {Lagrangian} observations},}\ }\href {\doibase 10.1016/j.pocean.2008.02.002}
  {\bibfield  {journal} {\bibinfo  {journal} {Progress in {Oceanography}}\
  }\textbf {\bibinfo {volume} {77}},\ \bibinfo {pages} {1--29} (\bibinfo {year}
  {2008})}\BibitemShut {NoStop}%
\bibitem [{\citenamefont {Newhouse}\ and\ \citenamefont
  {Pignataro}(1993)}]{Newhouse1993}%
  \BibitemOpen
  \bibfield  {author} {\bibinfo {author} {\bibfnamefont {S.}~\bibnamefont
  {Newhouse}}\ and\ \bibinfo {author} {\bibfnamefont {T.}~\bibnamefont
  {Pignataro}},\ }\bibfield  {title} {\enquote {\bibinfo {title} {On the
  estimation of topological entropy},}\ }\href {\doibase 10.1007/BF01048189}
  {\bibfield  {journal} {\bibinfo  {journal} {Journal of {Statistical}
  {Physics}}\ }\textbf {\bibinfo {volume} {72}},\ \bibinfo {pages} {1331--1351}
  (\bibinfo {year} {1993})}\BibitemShut {NoStop}%
\bibitem [{\citenamefont {Mariano}\ \emph {et~al.}(2002)\citenamefont
  {Mariano}, \citenamefont {Griffa}, \citenamefont {\"{O}zg\"{o}kmen},\ and\
  \citenamefont {Zambianchi}}]{Mariano2002}%
  \BibitemOpen
  \bibfield  {author} {\bibinfo {author} {\bibfnamefont {A.~J.}\ \bibnamefont
  {Mariano}}, \bibinfo {author} {\bibfnamefont {A.}~\bibnamefont {Griffa}},
  \bibinfo {author} {\bibfnamefont {T.~M.}\ \bibnamefont {\"{O}zg\"{o}kmen}}, \
  and\ \bibinfo {author} {\bibfnamefont {E.}~\bibnamefont {Zambianchi}},\
  }\bibfield  {title} {\enquote {\bibinfo {title} {Lagrangian {Analysis} and
  {Predictability} of {Coastal} and {Ocean} {Dynamics} 2000},}\ }\href
  {\doibase 10.1175/1520-0426(2002)019<1114:LAAPOC>2.0.CO;2} {\bibfield
  {journal} {\bibinfo  {journal} {Journal of {Atmospheric} and {Oceanic}
  {Technology}}\ }\textbf {\bibinfo {volume} {19}},\ \bibinfo {pages}
  {1114--1126} (\bibinfo {year} {2002})}\BibitemShut {NoStop}%
\bibitem [{\citenamefont {Puckett}\ \emph {et~al.}(2012)\citenamefont
  {Puckett}, \citenamefont {Lechenault}, \citenamefont {Daniels},\ and\
  \citenamefont {Thiffeault}}]{Puckett2012}%
  \BibitemOpen
  \bibfield  {author} {\bibinfo {author} {\bibfnamefont {J.~G.}\ \bibnamefont
  {Puckett}}, \bibinfo {author} {\bibfnamefont {F.}~\bibnamefont {Lechenault}},
  \bibinfo {author} {\bibfnamefont {K.~E.}\ \bibnamefont {Daniels}}, \ and\
  \bibinfo {author} {\bibfnamefont {J.-L.}\ \bibnamefont {Thiffeault}},\
  }\bibfield  {title} {\enquote {\bibinfo {title} {Trajectory entanglement in
  dense granular materials},}\ }\href {\doibase
  10.1088/1742-5468/2012/06/P06008} {\bibfield  {journal} {\bibinfo  {journal}
  {Journal of {Statistical} {Mechanics}: {Theory} and {Experiment}}\ }\textbf
  {\bibinfo {volume} {2012}},\ \bibinfo {pages} {P06008} (\bibinfo {year}
  {2012})}\BibitemShut {NoStop}%
\bibitem [{\citenamefont {Ali}(2013)}]{Ali2013}%
  \BibitemOpen
  \bibfield  {author} {\bibinfo {author} {\bibfnamefont {S.}~\bibnamefont
  {Ali}},\ }\bibfield  {title} {\enquote {\bibinfo {title} {Measuring {Flow}
  {Complexity} in {Videos}},}\ }in\ \href {\doibase 10.1109/ICCV.2013.140}
  {\emph {\bibinfo {booktitle} {2013 {IEEE} {International} {Conference} on
  {Computer} {Vision} ({ICCV})}}}\ (\bibinfo {year} {2013})\ pp.\ \bibinfo
  {pages} {1097--1104}\BibitemShut {NoStop}%
\bibitem [{\citenamefont {Caussin}\ and\ \citenamefont
  {Bartolo}(2015)}]{Caussin2015}%
  \BibitemOpen
  \bibfield  {author} {\bibinfo {author} {\bibfnamefont {J.-B.}\ \bibnamefont
  {Caussin}}\ and\ \bibinfo {author} {\bibfnamefont {D.}~\bibnamefont
  {Bartolo}},\ }\bibfield  {title} {\enquote {\bibinfo {title} {Braiding a
  flock: winding statistics of interacting flying spins},}\ }\href
  {http://arxiv.org/abs/1501.07879} {\bibfield  {journal} {\bibinfo  {journal}
  {{arXiv}:1501.07879 {[}cond-mat, physics:physics{]}}\ } (\bibinfo {year}
  {2015})}\BibitemShut {NoStop}%
\bibitem [{\citenamefont {Topaz}, \citenamefont {Ziegelmeier},\ and\
  \citenamefont {Halverson}(2014)}]{Topaz2014}%
  \BibitemOpen
  \bibfield  {author} {\bibinfo {author} {\bibfnamefont {C.~M.}\ \bibnamefont
  {Topaz}}, \bibinfo {author} {\bibfnamefont {L.}~\bibnamefont {Ziegelmeier}},
  \ and\ \bibinfo {author} {\bibfnamefont {T.}~\bibnamefont {Halverson}},\
  }\bibfield  {title} {\enquote {\bibinfo {title} {Topological {Data}
  {Analysis} of {Biological} {Aggregation} {Models}},}\ }\href
  {http://arxiv.org/abs/1412.6430} {\bibfield  {journal} {\bibinfo  {journal}
  {{arXiv}:1412.6430 {[}nlin, q-bio{]}}\ } (\bibinfo {year}
  {2014})}\BibitemShut {NoStop}%
\bibitem [{\citenamefont {Boyland}, \citenamefont {Aref},\ and\ \citenamefont
  {Stremler}(2000)}]{Boyland2000}%
  \BibitemOpen
  \bibfield  {author} {\bibinfo {author} {\bibfnamefont {P.~L.}\ \bibnamefont
  {Boyland}}, \bibinfo {author} {\bibfnamefont {H.}~\bibnamefont {Aref}}, \
  and\ \bibinfo {author} {\bibfnamefont {M.~A.}\ \bibnamefont {Stremler}},\
  }\bibfield  {title} {\enquote {\bibinfo {title} {Topological fluid mechanics
  of stirring},}\ }\href {\doibase 10.1017/S0022112099007107} {\bibfield
  {journal} {\bibinfo  {journal} {Journal of {Fluid} {Mechanics}}\ }\textbf
  {\bibinfo {volume} {403}},\ \bibinfo {pages} {277--304} (\bibinfo {year}
  {2000})}\BibitemShut {NoStop}%
\bibitem [{\citenamefont {Thiffeault}(2005)}]{Thiffeault2005}%
  \BibitemOpen
  \bibfield  {author} {\bibinfo {author} {\bibfnamefont {J.-L.}\ \bibnamefont
  {Thiffeault}},\ }\bibfield  {title} {\enquote {\bibinfo {title} {Measuring
  {Topological} {Chaos}},}\ }\href {\doibase 10.1103/PhysRevLett.94.084502}
  {\bibfield  {journal} {\bibinfo  {journal} {Physical {Review} {Letters}}\
  }\textbf {\bibinfo {volume} {94}},\ \bibinfo {pages} {084502} (\bibinfo
  {year} {2005})}\BibitemShut {NoStop}%
\bibitem [{\citenamefont {Thiffeault}(2010)}]{Thiffeault2010}%
  \BibitemOpen
  \bibfield  {author} {\bibinfo {author} {\bibfnamefont {J.-L.}\ \bibnamefont
  {Thiffeault}},\ }\bibfield  {title} {\enquote {\bibinfo {title} {Braids of
  entangled particle trajectories},}\ }\href {\doibase 10.1063/1.3262494}
  {\bibfield  {journal} {\bibinfo  {journal} {Chaos: {An} {Interdisciplinary}
  {Journal} of {Nonlinear} {Science}}\ }\textbf {\bibinfo {volume} {20}},\
  \bibinfo {pages} {017516--017514} (\bibinfo {year} {2010})}\BibitemShut
  {NoStop}%
\bibitem [{\citenamefont {Boyland}(1994)}]{Boyland1994}%
  \BibitemOpen
  \bibfield  {author} {\bibinfo {author} {\bibfnamefont {P.}~\bibnamefont
  {Boyland}},\ }\bibfield  {title} {\enquote {\bibinfo {title} {Topological
  methods in surface dynamics},}\ }\href {\doibase
  10.1016/0166-8641(94)00147-2} {\bibfield  {journal} {\bibinfo  {journal}
  {Topology and its {Applications}}\ }\textbf {\bibinfo {volume} {58}},\
  \bibinfo {pages} {223--298} (\bibinfo {year} {1994})}\BibitemShut {NoStop}%
\bibitem [{\citenamefont {Finn}, \citenamefont {Thiffeault},\ and\
  \citenamefont {Gouillart}(2006)}]{Finn2006}%
  \BibitemOpen
  \bibfield  {author} {\bibinfo {author} {\bibfnamefont {M.~D.}\ \bibnamefont
  {Finn}}, \bibinfo {author} {\bibfnamefont {J.-L.}\ \bibnamefont
  {Thiffeault}}, \ and\ \bibinfo {author} {\bibfnamefont {E.}~\bibnamefont
  {Gouillart}},\ }\bibfield  {title} {\enquote {\bibinfo {title} {Topological
  chaos in spatially periodic mixers},}\ }\href {\doibase
  10.1016/j.physd.2006.07.018} {\bibfield  {journal} {\bibinfo  {journal}
  {Physica {D}: {Nonlinear} {Phenomena}}\ }\textbf {\bibinfo {volume} {221}},\
  \bibinfo {pages} {92--100} (\bibinfo {year} {2006})}\BibitemShut {NoStop}%
\bibitem [{\citenamefont {Gouillart}, \citenamefont {Thiffeault},\ and\
  \citenamefont {Finn}(2006)}]{Gouillart2006}%
  \BibitemOpen
  \bibfield  {author} {\bibinfo {author} {\bibfnamefont {E.}~\bibnamefont
  {Gouillart}}, \bibinfo {author} {\bibfnamefont {J.-L.}\ \bibnamefont
  {Thiffeault}}, \ and\ \bibinfo {author} {\bibfnamefont {M.~D.}\ \bibnamefont
  {Finn}},\ }\bibfield  {title} {\enquote {\bibinfo {title} {Topological mixing
  with ghost rods},}\ }\href {\doibase 10.1103/PhysRevE.73.036311} {\bibfield
  {journal} {\bibinfo  {journal} {Physical {Review} {E}. {Statistical},
  {Nonlinear}, and {Soft} {Matter} {Physics}}\ }\textbf {\bibinfo {volume}
  {73}},\ \bibinfo {pages} {036311--036318} (\bibinfo {year}
  {2006})}\BibitemShut {NoStop}%
\bibitem [{\citenamefont {Thiffeault}\ and\ \citenamefont
  {Finn}(2006)}]{Thiffeault2006}%
  \BibitemOpen
  \bibfield  {author} {\bibinfo {author} {\bibfnamefont {J.-L.}\ \bibnamefont
  {Thiffeault}}\ and\ \bibinfo {author} {\bibfnamefont {M.~D.}\ \bibnamefont
  {Finn}},\ }\bibfield  {title} {\enquote {\bibinfo {title} {Topology, braids
  and mixing in fluids},}\ }\href {\doibase 10.1098/rsta.2006.1899} {\bibfield
  {journal} {\bibinfo  {journal} {Philosophical {Transactions} of the {Royal}
  {Society} {A}: {Mathematical}, {Physical} and {Engineering} {Sciences}}\
  }\textbf {\bibinfo {volume} {364}},\ \bibinfo {pages} {3251--3266} (\bibinfo
  {year} {2006})}\BibitemShut {NoStop}%
\bibitem [{\citenamefont {Handel}(1985)}]{Handel1985}%
  \BibitemOpen
  \bibfield  {author} {\bibinfo {author} {\bibfnamefont {M.}~\bibnamefont
  {Handel}},\ }\bibfield  {title} {\enquote {\bibinfo {title} {Global shadowing
  of pseudo-{Anosov} homeomorphisms},}\ }\href {\doibase
  10.1017/S0143385700003011} {\bibfield  {journal} {\bibinfo  {journal}
  {Ergodic {Theory} and {Dynamical} {Systems}}\ }\textbf {\bibinfo {volume}
  {5}},\ \bibinfo {pages} {373--377} (\bibinfo {year} {1985})}\BibitemShut
  {NoStop}%
\bibitem [{\citenamefont {Tumasz}\ and\ \citenamefont
  {Thiffeault}(2012)}]{Tumasz2012}%
  \BibitemOpen
  \bibfield  {author} {\bibinfo {author} {\bibfnamefont {S.}~\bibnamefont
  {Tumasz}}\ and\ \bibinfo {author} {\bibfnamefont {J.-L.}\ \bibnamefont
  {Thiffeault}},\ }\bibfield  {title} {\enquote {\bibinfo {title} {Topological
  {Entropy} and {Secondary} {Folding}},}\ }\href {\doibase
  10.1007/s00332-012-9159-9} {\bibfield  {journal} {\bibinfo  {journal}
  {Journal of {Nonlinear} {Science}}\ }\textbf {\bibinfo {volume} {23}},\
  \bibinfo {pages} {511--524} (\bibinfo {year} {2012})}\BibitemShut {NoStop}%
\bibitem [{\citenamefont {Tumasz}(2012)}]{Tumasz2012a}%
  \BibitemOpen
  \bibfield  {author} {\bibinfo {author} {\bibfnamefont {S.~E.}\ \bibnamefont
  {Tumasz}},\ }\emph {\bibinfo {title} {Topological stirring}},\ \href
  {http://gradworks.umi.com/35/46/3546864.html} {Ph.D. thesis},\ \bibinfo
  {school} {University of {Wisconsin}, {Madison}} (\bibinfo {year}
  {2012})\BibitemShut {NoStop}%
\bibitem [{\citenamefont {Tumasz}\ and\ \citenamefont
  {Thiffeault}(2013)}]{Tumasz2013}%
  \BibitemOpen
  \bibfield  {author} {\bibinfo {author} {\bibfnamefont {S.~E.}\ \bibnamefont
  {Tumasz}}\ and\ \bibinfo {author} {\bibfnamefont {J.-L.}\ \bibnamefont
  {Thiffeault}},\ }\bibfield  {title} {\enquote {\bibinfo {title} {Estimating
  {Topological} {Entropy} from the {Motion} of {Stirring} {Rods}},}\ }\href
  {\doibase 10.1016/j.piutam.2013.03.014} {\bibfield  {journal} {\bibinfo
  {journal} {Procedia {IUTAM}}\ }\textbf {\bibinfo {volume} {7}},\ \bibinfo
  {pages} {117--126} (\bibinfo {year} {2013})}\BibitemShut {NoStop}%
\bibitem [{\citenamefont {Allshouse}\ and\ \citenamefont
  {Thiffeault}(2012)}]{Allshouse2012}%
  \BibitemOpen
  \bibfield  {author} {\bibinfo {author} {\bibfnamefont {M.~R.}\ \bibnamefont
  {Allshouse}}\ and\ \bibinfo {author} {\bibfnamefont {J.-L.}\ \bibnamefont
  {Thiffeault}},\ }\bibfield  {title} {\enquote {\bibinfo {title} {Detecting
  coherent structures using braids},}\ }\href {\doibase
  10.1016/j.physd.2011.10.002} {\bibfield  {journal} {\bibinfo  {journal}
  {Physica {D}. {Nonlinear} {Phenomena}}\ ,\ \bibinfo {pages} {95--105}}
  (\bibinfo {year} {2012})}\BibitemShut {NoStop}%
\bibitem [{\citenamefont {Thiffeault}\ and\ \citenamefont
  {Budi\v{s}i\'{c}}(2014)}]{Thiffeault2014v3}%
  \BibitemOpen
  \bibfield  {author} {\bibinfo {author} {\bibfnamefont {J.-L.}\ \bibnamefont
  {Thiffeault}}\ and\ \bibinfo {author} {\bibfnamefont {M.}~\bibnamefont
  {Budi\v{s}i\'{c}}},\ }\bibfield  {title} {\enquote {\bibinfo {title}
  {Braidlab: {A} {Software} {Package} for {Braids} and {Loops} (v.3.1)},}\
  }\href {http://arxiv.org/abs/1410.0849v3} {\bibfield  {journal} {\bibinfo
  {journal} {{arXiv}:1410.0849v3 {[}math{]}}\ } (\bibinfo {year}
  {2014})}\BibitemShut {NoStop}%
\bibitem [{\citenamefont {Dynnikov}(2002)}]{Dynnikov2002}%
  \BibitemOpen
  \bibfield  {author} {\bibinfo {author} {\bibfnamefont {I.~A.}\ \bibnamefont
  {Dynnikov}},\ }\bibfield  {title} {\enquote {\bibinfo {title} {On a
  {Yang}-{Baxter} map and the {Dehornoy} ordering},}\ }\href {\doibase
  10.1070/RM2002v057n03ABEH000519} {\bibfield  {journal} {\bibinfo  {journal}
  {Russian {Mathematical} {Surveys}}\ }\textbf {\bibinfo {volume} {57}},\
  \bibinfo {pages} {592} (\bibinfo {year} {2002})}\BibitemShut {NoStop}%
\bibitem [{\citenamefont {Hall}\ and\ \citenamefont
  {Yurtta\c{s}}(2009)}]{Hall2009}%
  \BibitemOpen
  \bibfield  {author} {\bibinfo {author} {\bibfnamefont {T.}~\bibnamefont
  {Hall}}\ and\ \bibinfo {author} {\bibfnamefont {S.~O.}\ \bibnamefont
  {Yurtta\c{s}}},\ }\bibfield  {title} {\enquote {\bibinfo {title} {On the
  topological entropy of families of braids},}\ }\href {\doibase
  10.1016/j.topol.2009.01.005} {\bibfield  {journal} {\bibinfo  {journal}
  {Topology and its {Applications}}\ }\textbf {\bibinfo {volume} {156}},\
  \bibinfo {pages} {1554--1564} (\bibinfo {year} {2009})}\BibitemShut {NoStop}%
\bibitem [{Note1()}]{Note1}%
  \BibitemOpen
  \bibinfo {note} {The \protect \emph {minimum word length} is much more
  difficult to compute (see ref.~\protect \onlinecite {Paterson1991,
  Bangert2002}). Here we simply count the number of crossings and do not
  attempt to simplify the braid.}\BibitemShut {Stop}%
\bibitem [{\citenamefont {Fathi}, \citenamefont {Laudenbach},\ and\
  \citenamefont {Po\'{e}naru}(1979)}]{Fathi1979}%
  \BibitemOpen
  \bibfield  {author} {\bibinfo {author} {\bibfnamefont {A.}~\bibnamefont
  {Fathi}}, \bibinfo {author} {\bibfnamefont {F.}~\bibnamefont {Laudenbach}}, \
  and\ \bibinfo {author} {\bibfnamefont {V.}~\bibnamefont {Po\'{e}naru}},\
  }\href {http://www.ams.org/mathscinet-getitem?mr=568308} {\emph {\bibinfo
  {title} {Travaux de {Thurston} sur les surfaces}}},\ \bibinfo {series}
  {Ast\'{e}risque}, Vol.~\bibinfo {volume} {66}\ (\bibinfo  {publisher}
  {Soci\'{e}t\'{e} {Math}\'{e}matique de {France}, {Paris}},\ \bibinfo {year}
  {1979})\BibitemShut {NoStop}%
\bibitem [{\citenamefont {Birman}(1975)}]{Birman1975}%
  \BibitemOpen
  \bibfield  {author} {\bibinfo {author} {\bibfnamefont {J.~S.}\ \bibnamefont
  {Birman}},\ }\href@noop {} {\emph {\bibinfo {title} {Braids, {Links} and
  {Mapping} {Class} {Groups}}}},\ \bibinfo {series} {Annals of {Mathematics}
  {Studies}}\ No.~\bibinfo {number} {82}\ (\bibinfo  {publisher} {Princeton
  {University} {Press}},\ \bibinfo {address} {Princeton, {NJ}},\ \bibinfo
  {year} {1975})\BibitemShut {NoStop}%
\bibitem [{\citenamefont {Farb}\ and\ \citenamefont
  {Margalit}(2012)}]{Farb2012}%
  \BibitemOpen
  \bibfield  {author} {\bibinfo {author} {\bibfnamefont {B.}~\bibnamefont
  {Farb}}\ and\ \bibinfo {author} {\bibfnamefont {D.}~\bibnamefont
  {Margalit}},\ }\href {http://www.ams.org/mathscinet-getitem?mr=2850125}
  {\emph {\bibinfo {title} {A primer on mapping class groups}}},\ \bibinfo
  {series} {Princeton {Mathematical} {Series}}, Vol.~\bibinfo {volume} {49}\
  (\bibinfo  {publisher} {Princeton {University} {Press}, {Princeton}, {NJ}},\
  \bibinfo {year} {2012})\BibitemShut {NoStop}%
\bibitem [{\citenamefont {Franks}\ and\ \citenamefont
  {Handel}(1988)}]{Franks1988}%
  \BibitemOpen
  \bibfield  {author} {\bibinfo {author} {\bibfnamefont {J.~M.}\ \bibnamefont
  {Franks}}\ and\ \bibinfo {author} {\bibfnamefont {M.}~\bibnamefont
  {Handel}},\ }\bibfield  {title} {\enquote {\bibinfo {title} {Entropy and
  exponential growth of \${\textbackslash}pi\_1\$ in dimension two},}\ }\href
  {\doibase 10.2307/2047259} {\bibfield  {journal} {\bibinfo  {journal}
  {Proceedings of the {American} {Mathematical} {Society}}\ }\textbf {\bibinfo
  {volume} {102}},\ \bibinfo {pages} {753--760} (\bibinfo {year}
  {1988})}\BibitemShut {NoStop}%
\bibitem [{\citenamefont {Newhouse}(1988)}]{Newhouse1988}%
  \BibitemOpen
  \bibfield  {author} {\bibinfo {author} {\bibfnamefont {S.~E.}\ \bibnamefont
  {Newhouse}},\ }\bibfield  {title} {\enquote {\bibinfo {title} {Entropy and
  volume},}\ }\href {\doibase 10.1017/S0143385700009469} {\bibfield  {journal}
  {\bibinfo  {journal} {Ergodic {Theory} and {Dynamical} {Systems}}\ }\textbf
  {\bibinfo {volume} {8}},\ \bibinfo {pages} {283--299} (\bibinfo {year}
  {1988})}\BibitemShut {NoStop}%
\bibitem [{\citenamefont {Moussafir}(2006)}]{Moussafir2006}%
  \BibitemOpen
  \bibfield  {author} {\bibinfo {author} {\bibfnamefont {J.-O.}\ \bibnamefont
  {Moussafir}},\ }\bibfield  {title} {\enquote {\bibinfo {title} {On computing
  the entropy of braids},}\ }\href {\doibase 10.1007/s11853-007-0004-x}
  {\bibfield  {journal} {\bibinfo  {journal} {Functional {Analysis} and {Other}
  {Mathematics}}\ }\textbf {\bibinfo {volume} {1}},\ \bibinfo {pages} {37--46}
  (\bibinfo {year} {2006})}\BibitemShut {NoStop}%
\bibitem [{Note2()}]{Note2}%
  \BibitemOpen
  \bibinfo {note} {See Section~5 of the \protect \texttt {braidlab}
  guide~\protect \rev@citealpnum {Thiffeault2014v3} for a consequence of
  removing strings from braids of non-periodic trajectories.}\BibitemShut
  {Stop}%
\bibitem [{\citenamefont {Dynnikov}\ and\ \citenamefont
  {Wiest}(2007)}]{Dynnikov2007}%
  \BibitemOpen
  \bibfield  {author} {\bibinfo {author} {\bibfnamefont {I.}~\bibnamefont
  {Dynnikov}}\ and\ \bibinfo {author} {\bibfnamefont {B.}~\bibnamefont
  {Wiest}},\ }\bibfield  {title} {\enquote {\bibinfo {title} {On the complexity
  of braids},}\ }\href {\doibase 10.4171/JEMS/98} {\bibfield  {journal}
  {\bibinfo  {journal} {Journal of the {European} {Mathematical} {Society}
  ({JEMS})}\ }\textbf {\bibinfo {volume} {9}},\ \bibinfo {pages} {801--840}
  (\bibinfo {year} {2007})}\BibitemShut {NoStop}%
\bibitem [{Note4()}]{Note4}%
  \BibitemOpen
  \bibinfo {note} {Relative Standard Deviation (RSD) is the standard deviation
  divided by the mean.}\BibitemShut {Stop}%
\bibitem [{\citenamefont {Hennion}(1997)}]{Hennion1997}%
  \BibitemOpen
  \bibfield  {author} {\bibinfo {author} {\bibfnamefont {H.}~\bibnamefont
  {Hennion}},\ }\bibfield  {title} {\enquote {\bibinfo {title} {Limit theorems
  for products of positive random matrices},}\ }\href {\doibase
  10.1214/aop/1023481103} {\bibfield  {journal} {\bibinfo  {journal} {The
  {Annals} of {Probability}}\ }\textbf {\bibinfo {volume} {25}},\ \bibinfo
  {pages} {1545--1587} (\bibinfo {year} {1997})}\BibitemShut {NoStop}%
\bibitem [{\citenamefont {Young}(1998)}]{Young1998}%
  \BibitemOpen
  \bibfield  {author} {\bibinfo {author} {\bibfnamefont {L.-S.}\ \bibnamefont
  {Young}},\ }\bibfield  {title} {\enquote {\bibinfo {title} {Statistical
  {Properties} of {Dynamical} {Systems} with {Some} {Hyperbolicity}},}\ }\href
  {\doibase 10.2307/120960} {\bibfield  {journal} {\bibinfo  {journal} {Annals
  of {Mathematics}}\ }\bibinfo {series} {Second {Series}},\ \textbf {\bibinfo
  {volume} {147}},\ \bibinfo {pages} {585--650} (\bibinfo {year}
  {1998})}\BibitemShut {NoStop}%
\bibitem [{\citenamefont {Finn}\ and\ \citenamefont
  {Thiffeault}(2007)}]{Finn2007}%
  \BibitemOpen
  \bibfield  {author} {\bibinfo {author} {\bibfnamefont {M.~D.}\ \bibnamefont
  {Finn}}\ and\ \bibinfo {author} {\bibfnamefont {J.-L.}\ \bibnamefont
  {Thiffeault}},\ }\bibfield  {title} {\enquote {\bibinfo {title} {Topological
  entropy of braids on the torus},}\ }\href {\doibase 10.1137/060659636}
  {\bibfield  {journal} {\bibinfo  {journal} {{SIAM} {Journal} on {Applied}
  {Dynamical} {Systems}}\ }\textbf {\bibinfo {volume} {6}},\ \bibinfo {pages}
  {79--98 (electronic)} (\bibinfo {year} {2007})}\BibitemShut {NoStop}%
\bibitem [{\citenamefont {Clauset}, \citenamefont {Shalizi},\ and\
  \citenamefont {Newman}(2009)}]{Clauset2009}%
  \BibitemOpen
  \bibfield  {author} {\bibinfo {author} {\bibfnamefont {A.}~\bibnamefont
  {Clauset}}, \bibinfo {author} {\bibfnamefont {C.~R.}\ \bibnamefont
  {Shalizi}}, \ and\ \bibinfo {author} {\bibfnamefont {M.~E.~J.}\ \bibnamefont
  {Newman}},\ }\bibfield  {title} {\enquote {\bibinfo {title} {Power-law
  distributions in empirical data},}\ }\href {\doibase 10.1137/070710111}
  {\bibfield  {journal} {\bibinfo  {journal} {{SIAM} {Review}}\ }\textbf
  {\bibinfo {volume} {51}},\ \bibinfo {pages} {661--703} (\bibinfo {year}
  {2009})}\BibitemShut {NoStop}%
\bibitem [{\citenamefont {Mitzenmacher}(2004)}]{Mitzenmacher2004}%
  \BibitemOpen
  \bibfield  {author} {\bibinfo {author} {\bibfnamefont {M.}~\bibnamefont
  {Mitzenmacher}},\ }\bibfield  {title} {\enquote {\bibinfo {title} {A brief
  history of generative models for power law and lognormal distributions},}\
  }\href {\doibase 10.1080/15427951.2004.10129088} {\bibfield  {journal}
  {\bibinfo  {journal} {Internet {Mathematics}}\ }\textbf {\bibinfo {volume}
  {1}},\ \bibinfo {pages} {226--251} (\bibinfo {year} {2004})}\BibitemShut
  {NoStop}%
\bibitem [{\citenamefont {Paterson}\ and\ \citenamefont
  {Razborov}(1991)}]{Paterson1991}%
  \BibitemOpen
  \bibfield  {author} {\bibinfo {author} {\bibfnamefont {M.~S.}\ \bibnamefont
  {Paterson}}\ and\ \bibinfo {author} {\bibfnamefont {A.~A.}\ \bibnamefont
  {Razborov}},\ }\bibfield  {title} {\enquote {\bibinfo {title} {The set of
  minimal braids is co-{NP}-complete},}\ }\href {\doibase
  10.1016/0196-6774(91)90011-M} {\bibfield  {journal} {\bibinfo  {journal}
  {Journal of {Algorithms}. {Cognition}, {Informatics} and {Logic}}\ }\textbf
  {\bibinfo {volume} {12}},\ \bibinfo {pages} {393--408} (\bibinfo {year}
  {1991})}\BibitemShut {NoStop}%
\bibitem [{\citenamefont {Bangert}, \citenamefont {Berger},\ and\ \citenamefont
  {Prandi}(2002)}]{Bangert2002}%
  \BibitemOpen
  \bibfield  {author} {\bibinfo {author} {\bibfnamefont {P.~D.}\ \bibnamefont
  {Bangert}}, \bibinfo {author} {\bibfnamefont {M.~A.}\ \bibnamefont {Berger}},
  \ and\ \bibinfo {author} {\bibfnamefont {R.}~\bibnamefont {Prandi}},\
  }\bibfield  {title} {\enquote {\bibinfo {title} {In search of minimal random
  braid configurations},}\ }\href {\doibase 10.1088/0305-4470/35/1/304}
  {\bibfield  {journal} {\bibinfo  {journal} {Journal of {Physics}. {A}.
  {Mathematical} and {General}}\ }\textbf {\bibinfo {volume} {35}},\ \bibinfo
  {pages} {43--59} (\bibinfo {year} {2002})}\BibitemShut {NoStop}%
\end{thebibliography}%

\end{document}